\documentclass{ws-ijmpc}
\usepackage{amsmath}
\usepackage{amssymb}
\usepackage{graphicx}
\usepackage{color}
\usepackage{nicefrac}
\usepackage{url,overcite}
\usepackage{subfigure}

\def\bib{B\kern-.05em{I}\kern-.025em{B}\kern-.08em}
\def\btex{B\kern-.05em{I}\kern-.025em{B}\kern-.08em\TeX}

\newcommand{\note}[1]{\emph{\textcolor{red}{}}}

\newcommand{\Msun}{{\ensuremath{\mathrm{M}_{\odot}}}}
\newcommand{\Zsun}{\ensuremath{Z_\odot}}

\newcommand{\kpc}{\ensuremath{\mathrm{kpc}}}

\newcommand{\Myr}{\ensuremath{\mathrm{Myr}}}
\newcommand{\K}{\ensuremath{\mathrm{K}}}

\newcommand{\eV}{\ensuremath{\mathrm{e\!V}}}

\newcommand{\gcc}{\ensuremath{\mathrm{g}\,\mathrm{cm}^{-3}}}
\newcommand{\gamad}{\ensuremath{\gamma_{\mathrm{ad}}}}

\newcommand{\HI}{H~I}
\newcommand{\HeI}{He~I}
\newcommand{\HII}{H~II}
\newcommand{\HeII}{He~II}

\newcommand{\Ni}{{\ensuremath{^{56}\mathrm{Ni}}}}
\newcommand{\Fe}{{\ensuremath{^{56}\mathrm{Fe}}}}
\newcommand{\jFe}{{\ensuremath{^{54}\mathrm{Fe}}}}
\newcommand{\iFe}{{\ensuremath{^{52}\mathrm{Fe}}}}
\newcommand{\Co}{{\ensuremath{^{56}\mathrm{Co}}}}
\newcommand{\He}{{\ensuremath{^{4} \mathrm{He}}}}
\newcommand{\Hea}{{\ensuremath{^{3} \mathrm{He}}}}
\newcommand{\Heb}{{\ensuremath{^{4} \mathrm{He}}}}
\newcommand{\Hy}{{\ensuremath{^{1} \mathrm{H}}} }
\newcommand{\Ox}{{\ensuremath{^{16}\mathrm{O}}}}
\newcommand{\Ti}{{\ensuremath{^{44}\mathrm{Ti}}}}
\newcommand{\Si}{{\ensuremath{^{28}\mathrm{Si}}}}
\newcommand{\Mg}{{\ensuremath{^{24}\mathrm{Mg}}}}
\newcommand{\Mgx}{{\ensuremath{^{23}\mathrm{Mg}}}}
\newcommand{\Mgy}{{\ensuremath{^{24}\mathrm{Mg}}}}
\newcommand{\Pxx}{{\ensuremath{^{31}\mathrm{P}}}}
\newcommand{\Cx}{{\ensuremath{^{12}\mathrm{C}}}}
\newcommand{\Be}{{\ensuremath{^{8}\mathrm{Be}}}}

\newcommand{\Cr}{{\ensuremath{^{48}\mathrm{Cr}}}}
\newcommand{\Ca}{{\ensuremath{^{40}\mathrm{Ca}}}}
\newcommand{\Na}{{\ensuremath{^{23}\mathrm{Na}}}}
\newcommand{\Ar}{{\ensuremath{^{36}\mathrm{Ar}}}}
\newcommand{\Sx}{{\ensuremath{^{32}\mathrm{S}}}}
\newcommand{\Nx}{{\ensuremath{^{14}\mathrm{N}}}}
\newcommand{\Ne}{{\ensuremath{^{20}\mathrm{Ne}}}}
\newcommand{\Ha}{{\ensuremath{^{} \mathrm{H}}} }
\newcommand{\Hm}{{\ensuremath{^{} \mathrm{H}_2}}}
\newcommand{\ex}{{\ensuremath{^{} \mathrm{e}^-}}}

\newcommand{\MS}{{\ensuremath{M_*}}}

\def\omegadot {{\dot\omega}}

\def\gb {{\bf g}}

\def\ub {{\bf u}}

\newcommand{\Cplusplus}{{\rmfamily C\raise.22ex\hbox{\small ++} }}

\newcommand{\xEq}[1]{eq:#1}

\newcommand{\lEq}[1]{\label{\xEq{#1}}}

\newcommand{\Eqref}[1]{{\ref{eq:#1}}}
\newcommand{\Eqff}[1]{{(\Eqref{#1})}}

\newcommand{\Eq}[1]{{Equation~\Eqff{#1}}}

\newcommand{\Ep}[1]{{\ensuremath{10^{#1}}}}

\newcommand{\cm}{{\ensuremath{\mathrm{cm}}}}
\newcommand{\km}{{\ensuremath{\mathrm{km}}}}
\newcommand{\gx}{{\ensuremath{\mathrm{g}}}}
\newcommand{\KT}{{\ensuremath{\mathrm{K}}}}

\newcommand{\usec}{{\ensuremath{\mathrm{sec}}}}
\newcommand{\pc}{{\ensuremath{\mathrm{pc}}}}
\newcommand{\erg}{{\ensuremath{\mathrm{erg}}}}

\newcommand{\Mpc}{{\ensuremath{\mathrm{Mpc}}}}
\newcommand{\GIFT}{\texttt{GIFT}}
\newcommand{\MA}{\texttt{MA28}}
\newcommand{\CASTRO}{\texttt{CASTRO}}
\newcommand{\KEPLER}{\texttt{KEPLER}}
\newcommand{\GADGET}{\texttt{GADGET}}
\newcommand{\VISIT}{\texttt{VISIT}}

\newcommand{\FLASH}{\texttt{FLASH}}
\newcommand{\MESA}{\texttt{MESA}}

\newcommand{\aj}{AJ}%
\newcommand{\araa}{ARA\&A}%
\newcommand{\apj}{ApJ}%
\newcommand{\apjl}{{ApJ}}%
\newcommand{\apjs}{{ApJS}}%
\newcommand{\apss}{{Ap\&SS}}%
\newcommand{\aap}{{A\&A}}%
\newcommand{\mnras}{{MNRAS}}%
\newcommand{\prd}{{Phys.~Rev.~D}}%
\newcommand{\pasp}{{PASP}}%
\newcommand{\ssr}{{Space~Sci.~Rev.}}%
\newcommand{\nat}{{Nature}}%
\newcommand{\nar}{{Nature}}%
\newcommand{\physrep}{{Phys.~Rep.}}%
\newcommand{\na}{{New Astronomy}}%

\newcommand{\sfrac}[2]{\mathchoice
  {\kern0em\raise.5ex\hbox{\the\scriptfont0 #1}\kern-.15em/
   \kern-.15em\lower.25ex\hbox{\the\scriptfont0 #2}}
  {\kern0em\raise.5ex\hbox{\the\scriptfont0 #1}\kern-.15em/
   \kern-.15em\lower.25ex\hbox{\the\scriptfont0 #2}}
  {\kern0em\raise.5ex\hbox{\the\scriptscriptfont0 #1}\kern-.2em/
   \kern-.15em\lower.25ex\hbox{\the\scriptscriptfont0 #2}}
  {#1\!/#2}}

\def\gb {{\bf g}}

\def\ub {{\bf u}}

\def\omegadot {\dot\omega}

\newcommand{\be}{\begin{equation}}
\newcommand{\ee}{\end{equation}}
\newcommand{\bea}{\begin{eqnarray}}
\newcommand{\eea}{\end{eqnarray}}
\newcommand{\bc}{\begin{center}}
\newcommand{\ec}{\end{center}}

\renewcommand{\vec}[1]{\textit{\textbf{#1}}}

\newcommand{\cc}{\ensuremath{\mathrm{cm}^{-3}}}

\begin{document}

\markboth{Chen}
{SUPERNOVAE AT THE COSMIC DAWN}

\catchline{}{}{}{}{}

\title{SUPERNOVAE AT THE COSMIC DAWN  \footnote{}
}

\author{KE-JUNG CHEN }

\address{Department of Astronomy \& Astrophysics, University of California, 1156 High St.\\ 
Santa Cruz, California 95064, USA \\ 
School of Physics and Astronomy, University of Minnesota, 116 Church St. \\
Minneapolis, Minnesota 55455, USA \\ 
kchen@ucolick.org
}





\maketitle

\begin{history}
\received{Day Month Year}
\revised{Day Month Year}
\end{history}

\begin{abstract}
Modern cosmological simulations predict that the first generation of stars formed with a mass scale
around $100\,\Msun$ about $300-400$ million years after the Big Bang. When the first 
stars reached the end of their lives, many of them might have died as energetic supernovae  that 
could have significantly affected the early Universe via injecting large amounts of energy and metals 
into the primordial intergalactic medium. In this paper, we review  the current models of the first 
supernovae by discussing on the relevant background physics, computational methods, and the latest results.

\end{abstract}

\ccode{PACS Nos.: 97.60.Bw, 98.80.Bp, 98.80.Ft}

\section{Introduction}
One of the frontiers in modern cosmology is understanding the end of the cosmic dark ages, 
when the first luminous objects (e.g., stars, supernovae (SNe), and galaxies) reshaped the 
primordial Universe into the current Universe. The advancement of 
supercomputing power in the last decade has allowed us to 
start investigating the formation of the first stars by modeling the relevant physical processes. 
The results of the first star formation suggested that these stars could have been very massive, 
having a typical mass scale of about $100$ solar masses (\Msun). Some of them might 
have died as energetic SN explosions. These first SNe could dump considerable energy and 
spread the previously-forged elements to the inter-galactic medium (IGM) that significantly 
impacted later star formation. The forthcoming observatories will soon probe these first SNe; 
therefore, it is timely that we review the current theoretical models about the first SNe. In this review, we present a brief overview of modern cosmology in \S~\ref{earlyU} and the physics of 
the first  star formation and its stellar evolution in \S~\ref{firststar}.  We then discuss the computational 
approaches for simulating the first SNe  in \S~\ref{sn_castro}.  
We discuss the explosion mechanics of the first SNe by presenting some of latest results in \S~\ref{sn_result} .  The yields and energetics of these first SNe might affect the early Universe, which then transformed into
the present Universe. We introduce the computational approaches for feedback simulations of the first stars and SNe in \S~\ref{feed_gadget} and present the results in \S~\ref{feed_result}.  Finally, we give a summary and perspective in \S~\ref{summary}.

\section{The Early Universe}
\label{earlyU}
The creation and evolution of the Universe has been one of the most fascinating subjects 
in modern cosmology. It is proper to provide the background of the early Universe, which hatched
the first stars and supernovae, which are the major topics of this review. 
This section provides a brief overview of modern cosmology. 
There are many excellent reviews about the early Universe;  we list only some of them 
for readers interested in having a more comprehensive understanding of 
modern cosmology. The recommended entry-level textbooks about the early Universe are 
[\refcite{liddle2003}] for undergraduate students and [\refcite{peacock1999}] for graduate students.
For more specific studies, [\refcite{kolb1990}] provides a comprehensive introduction to the Inflationary 
model and the Big Bang Nucleosynthesis. [\refcite{dodelson2003}] discusses the quantum fluctuation from 
Inflation and how it was seeded as initial perturbations for the large scale structure formation. Those 
who are interested in the dynamics and evolution of the Universe can consult two classic 
books: [\refcite{peebles1980,peebles1993}]. 

Our Universe is believed to have been born from the Big Bang at the time when the density and temperature 
of the Universe were infinite. At the beginning of the Big Bang, all fundamental physical forces---such as
gravitational, electro-magnetic, strong, and weak forces---were united. Due to the rapid expansion of the Universe,
the temperature dropped quickly, and the fundamental forces became separated. At about $10^{-36}\,\sec$  
after the Big Bang commenced, the Universe went through a very short and rapid expansion called Inflation \cite{guth1981,linde1982}{}. Inflation seeded the quantum fluctuations into space-time. These fluctuations later became the initial perturbations of the Universe, which led to the formation of large scale structures. 
A few minutes later,  atomic nuclei could start to form.  Then protons  and neutrons began to combine into atomic nuclei: helium (24\% in mass), hydrogen (76 \% in mass), and a trace amount of lithium. The Big Bang Nucleosynthesis lasted only until the temperatures and densities of baryons 
became too low for further nucleosynthesis, which was about a few minutes. The elements 
necessary for life, such as carbon and oxygen, had not been made at this moment. 

About 300,000 years after the Big Bang, the temperature of Universe cooled below $10,000\,\K$. 
At that time, protons and electrons could recombine into neutral hydrogen. Without the opacity from free electrons, the photons decoupled 
from the matter and streamed freely. This radiation is called the cosmic microwave background radiation (CMB), and it was first detected by 
[\refcite{penzias1965}]. It fits perfectly with a black-body  temperature of about $2.73\,\K$. In 1992, the Cosmic Background Explorer (COBE) 
detected the anisotropy of the CMB, which shed the light of understanding on the structure formation of the early Universe. More recent results 
from the Wilkinson Microwave Anisotropy Probe (WMAP) helped to confirm inflationary cosmology and determined the cosmological parameters 
with an unprecedented precision. The success of the CMB observation confirmed that the Universe contains about 
$5\,\%$ of baryon, $25\,\%$ of cold dark matter (CDM), and $70\%$ of dark energy ($\Lambda$). The intrinsic properties of cold dark matter and dark energy remain poorly understood. Significant experimental effort has 
been made for studying the dark sectors of the Universe; promising progress should be made in the near future. 
Nevertheless, the Big Bang Nucleosynthesis, inflationary models, and $\Lambda$CDM form the foundation of 
modern cosmology. 
 
\begin{figure}[ht]
  \begin{center}
\includegraphics[width=\columnwidth]{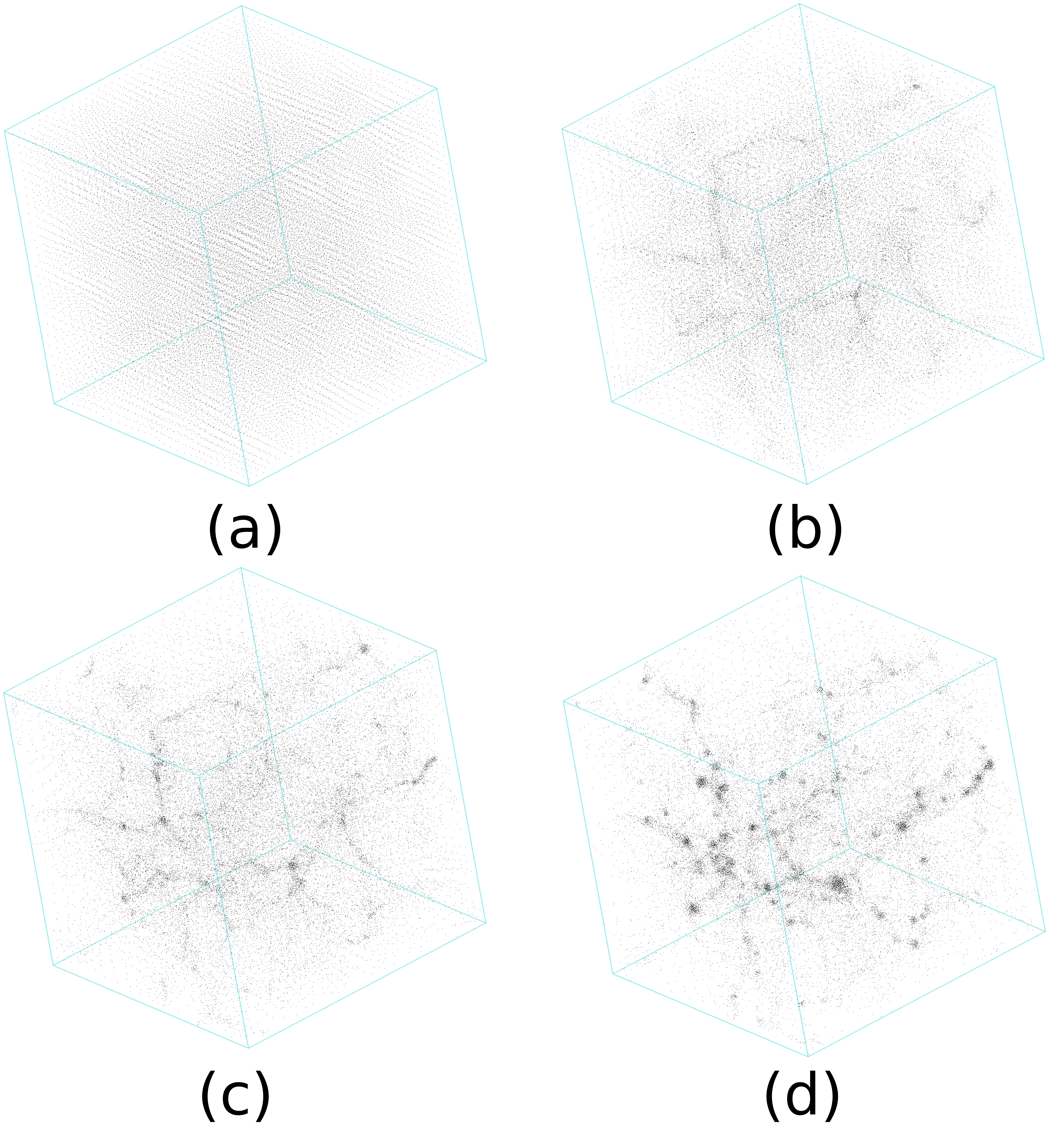}
\caption[]{ The formation of large scale structure of Universe: The black dots represent the 
dark matter particles. The evolution follows from panel (a) $\rightarrow$ (b) $\rightarrow$ (c) $\rightarrow$ (d). Cold dark matter particles only interact with each others through gravity and eventually form into a clustered structure. }
\label{LCDM_fig}
\end{center}
 \end{figure}

The initial perturbation seeded by inflation began evolving through gravity. In Figure~\ref{LCDM_fig}, we show the formation 
of a large scale structure from our cosmological simulation \cite{chen_phd} with \GADGET. This example consists of $128^3$ 
dark matter particles, and $128^3$ gas particles, following structure formation in a periodic box of size $50\,h^{-1}\,\Mpc^3$ 
in a $\Lambda$CDM Universe. The simulation begins at the redshift of $z\sim10$ and ends at $z\sim0$.  The initial 
distribution of particles was homogeneous and isotopic with a very tiny gaussian fluctuation. At the end, the dark matter 
particles (black dots) evolved into highly clustered structures hierarchically through gravity. 

\begin{figure}[ht]
\begin{center}
\includegraphics[width=\columnwidth]{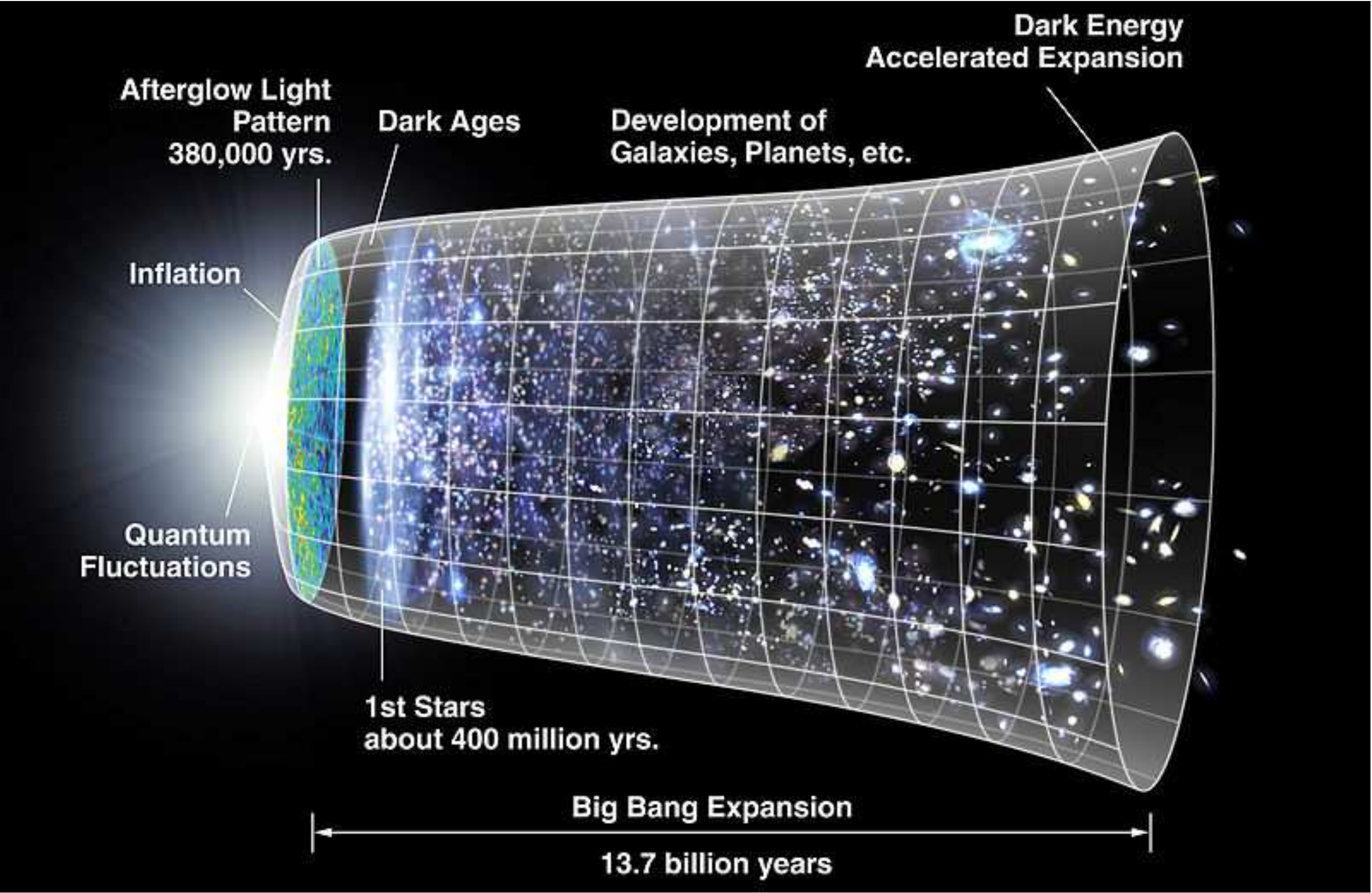}
\caption[]{Cosmic timeline:  The illustration shows the evolution of cosmic structure from the Big Bang. 
The first stars appeared about 400 million years after the Big Bang. Hereafter, the galaxies, stars, and planets
started to develop. Recent observations suggest that the expansion of the Universe is accelerating due to 
dark energy. 
(Credit: NASA/WMAP Science Team)\label{timeline}}
\end{center}
\end{figure}

There was no star when the CMB was emitted because the density of primordial gas was 
too low and could not condense to form stars. The Universe then entered the cosmic
dark ages when there was no light from stars. Several hundred million years after the 
Big Bang, the dark matter collapsed into minihalos with masses of $10^5-10^6\,\Msun$,
which would become the birth sites for the first stars because such halos could provide gravitational wells 
that retained the gas to form stars. The light from the first stars ended the dark ages, which had lasted for 
several hundred million years. In addition, the first stars started to forge the first metals that became the 
building blocks of later stars and galaxies. Thus, the first stars play a crucial role in the evolution of the Universe. Figure~\ref{timeline} shows a timeline of the Universe. 
The observable Universe spans about 13.7 billion years, starting with the Big Bang and quickly expanding
 during Inflation. After $380,000$ years, the CMB was emitted from the last scattering surface. Later, 
 the Universe entered the dark ages until the first stars were born. Hereafter, the planets, stars, and 
 galaxies started to form.

\section{The First Stars}
\label{firststar}
Formation of the first stars transformed the simple early Universe into a highly complicated 
one. The first stars made from the hydrogen and helium left from the Big Bang are called the 
Population~III (Pop~III) stars, which are ancestors of the current stars like our Sun. The study 
of the first stars has recently received increasing attention because the tools for this study have 
become available, including the forthcoming telescopes, which will probe the cosmic dark 
ages, and the advancement of modern supercomputers, which allow us to carry out more sophisticated 
simulations. In this section, we review the recent advancement of our understanding of the first star formation.

The $\Lambda$CDM model offers a fundamental theory for the large scale formation, 
suggesting that the cosmic structure formed in a hierarchical manner. The first stars must form 
along with the structural evolution of the Universe. The conditions for the star 
formation are that the cooling time scale of halos must be smaller than their dynamical time scale. 
According to [\refcite{bromm2004a}], the low-mass dark matter halos have a virial temperature of 
$\propto M^{2/3}(1+z)$,  where $M$ is the halo mass and $z$ is the redshift. Metal cooling was absent in the 
early Universe, and the cooling of gas occurred primarily through molecular hydrogen, H$_2$. The 
dominating H$_2$ formation goes through H $+\,e^-$ $\rightarrow$  H$^-+\,\gamma$ 
and  H$^-$ + H $\rightarrow$ H$_2$ $+\,e^-$. Sources of free electrons, $e^-$, come from the recombination 
or collision excitation of gas when dark matter halos merge. Pioneering  work\cite{tegmark1997, 
abel2002} suggests that the first star was born in the halos of $\sim 10^6\,\Msun$ at $z \sim 30$, 
which reach a H$_2$ fraction of $10^{-4}$. The size of Pop~III star-forming clouds is comparable 
to the virial radius of the halos, about 100~pc. The detailed shape of the cloud is determined by 
its angular momentum, which depends on the resolution of the simulations. Now 
there is no direct detection of Pop~III stars. Nevertheless,  the observation of present-day stars 
may provide us hints to study the Pop~III star formation. The present-day (Pop~I) stars are born 
inside a giant molecular cloud of about 100~pc, supported by the pressure of turbulence flow 
or magnetic field.  About $1,000\,-\,1,000,000$ stars usually form inside the cloud
 [\refcite{salpeter1955}], which suggests the observed initial mass function (IMF) of Pop~I stars to be
\begin{equation}
 N(M_*)=N_0\,M_*^{-2.35},
\end{equation}
where $N$ is the number of stars, $M_*$ is the stellar mass, and $N_0$
is a constant. The characteristic mass scale of the Salpeter IMF is about  $1\,\Msun$, which means 
most of the Pop~I stars form with a mass similar to that of our Sun.  It is extremely difficult to calculate the Pop~I 
IMF from first principles because present-day star formation involves magneto-hydrodynamics, turbulent flow, and complex chemistry. 
However, the initial conditions of the primordial Universe, such as the cosmological parameters, are better understood.  In addition,  the metal-free and
magnetic-free gas makes the simulation of Pop~III star formation more accessible.  To simulate the Pop~III star formation, we need 3D cosmological 
simulations of dark matter and gas, including cooling and chemistry for primordial gas. The initial conditions of simulations use the cosmological 
parameters from the CMB measurement. 

The key feature for cosmological simulation is handling a large dynamical range. Two popular setups for simulating 
the first star formation are mesh-based [\refcite{abel2000,abel2002}] and Lagrangian techniques [\refcite{bromm2002,bromm2009}]. 
The mesh-based technique usually employs the 
adaptive mesh refinement (AMR), which creates finer grids to resolve the structures of interests such as gas flow inside 
the dark matter halos. The other approach is called smoothed particle hydrodynamics (SPH), 
which uses particles to model the fluid elements. The mass distribution of particles is based on a kernel function.  
The results of AMR and SPH simulations both agree on the characteristics of the first star-forming cloud; temperature of $T_c\approx200\,\K$, and gas density of $n_c\approx10^4\,\cm^{-3}$. The $T_c$ is determined by H$_2$ cooling, which is the dominating coolant at that time. 
The lowest energy levels of H$_2$ are collisional excitation and subsequent rotational transitions with an energy gap of 
$\Delta E/k_{\rm B}\,\simeq\,512\,\K$.  Atomic hydrogen can cool down to several hundred K through collisions with H$_2$; 
$n_c$ is explained by the saturation of H$_2$ cooling: below $n_c$, the cooling rate is $\propto n^2$; 
above $n_c$, the cooling rate is $\propto n$.  Once the gas reaches the characteristic status, the cooling then becomes
inefficient and the gas cloud becomes a quasi-hydrostatic. The cloud eventually collapses when the its mass is larger than its 
Jeans mass\cite{bromm2004a}{},
\begin{equation}
 M_{J} = 700\bigg(\frac{T}{200\,{\rm K}}\bigg)^{3/2} \bigg(\frac{n}{10^4\,{\rm cm^{-3}}}\bigg)
 ^{-1/2}\quad \Msun.
\end{equation}
The Jeans mass is determined by the balance between the gravity and pressure of gas.  For the first star formation, the pressure is mainly 
from the thermal pressure of the gas. However, it is unclear whether the cloud forms into a single star or fragments into multiple stars. To answer this question, 
evolving the cloud to a higher density and following the subsequent accretion are required. The cloud mass at least sets up a maximum mass for the final 
stellar mass. But the exact mass of the stars is determined by the accretion history when the star forms.  [\refcite{bromm2004b}] suggested that the 
first stars can be very massive, having a typical mass of $100\,\Msun$ with a broad spectrum of mass distribution. 
 
\subsection{Stellar Evolution}
After the first star has formed, its core temperature increases due to Kelvin-Helmholtz contraction and eventually 
ignites hydrogen burning. In contrast to the present-day stars, there was no metal present inside the first stars. 
They first burn hydrogen into helium through p-p chains, then burn helium through the $3\alpha$ reaction. A detailed description 
of hydrogen burning can be found in [\refcite{prian2000}]. After the first carbon and oxygen have been made, the 
first stars can burn the hydrogen in a more effective way through the carbon-nitrogen-oxygen (CNO) cycle. Once stable hydrogen burning at the core of the star occurs, the first stars enter their main sequence. The lifespan of a star on the main sequence mainly depends on its initial mass and composition. The energy released from nuclear burning is used to power the luminosity of stars. 
Once the hydrogen is depleted, the star completes the main sequence and starts to burn helium as well 
as the resulting nuclei. In the following subsections, we introduce the advanced burning stages of stars before they die.

The luminosity of stars is powered by the nuclear fusion that occurs inside the stars. 
Light elements are synthesized into heavy elements, and the accompanying energy is released. 
We review the advanced burning stages based on [\refcite{kippen1990,arnett1996,prian2000,woosley2002}]. 
First, the helium burning consists of two steps, 
\begin{equation}
 \He\,+\,\He\,\rightarrow\,\Be, \quad
 \Be\,+\,\He\,\rightarrow\,\Cx. 
\end{equation}
The process is known as the $3\,\alpha$ reaction because three helium ($\alpha$) are involved.  It yields 
$5.8\,\times \,10^{17}\,\erg\,\gx^{-1}$. \Be{} determines the overall reaction rate, and its production 
is proportional to the square of the \He{} number density. So the energy generation rate is proportional to the density square. 
The formula of the energy generation rate of the $3\,\alpha$ reaction \cite{prian2000}  is
\begin{equation}
q_{3\alpha}\,\propto\,\rho^2 T^{40}.
\end{equation}
Some $\alpha$ capture reaction may occur, if sufficient amount of $\Cx$ are present. But at such a temperature, only
\begin{equation}
 \Cx\,+\,\He\,\rightarrow\,\Ox 
\end{equation}
is significant; other capture reaction rates are too low. So the major products of 
helium burning are carbon and oxygen, and the ratio of $\Cx/\Ox$ depends 
on temperature. After the helium burning,  the star starts to burn carbon and oxygen, which 
require higher temperatures to ignite. Carbon starts to burn when the temperatures reach 
$5\,\times\,10^8\,\KT$. There are several channels of carbon burning, 
\begin{eqnarray*}
   &&  \Mgx\quad + \quad \gamma \\
   &&  \Mgy \quad +\quad  n \\
\Cx\,+\,\Cx   &\longrightarrow&  \Na \quad\, + \quad  p \\
   && \Ne \,\quad +\quad  \alpha \\
   &&  \hphantom{0}\Ox \,\quad +\quad  2\alpha.   
\end{eqnarray*}
The overall energy generation is about  $5.2\,\times\,10^{17}\,\erg\,\gx^{-1}$. The process of oxygen burning ignites at a temperature of $10^9\,\K$. Similar to $\Cx$, there are several channels available: 

\begin{eqnarray*}
   &&  \Si \quad +\quad  \gamma \\
   &&  \Sx \,\,\quad  + \quad  n \\
\Ox\,+\,\Ox   &\longrightarrow&  \Pxx \,\quad  +\quad  p \\
   && \Si \,\quad + \quad \alpha \\
   && \Mg \!\!\quad + \quad 2\alpha.   
\end{eqnarray*}
The average energy released is about $4.8\,\times,\,10^{17}\,\erg\,\gx^{-1}$.
There is little interaction between carbon and oxygen for the intermediate 
temperature that ignites carbon burning because the carbon can quickly burn 
out by self interaction.  The light elements produced from carbon 
and oxygen burning are immediately captured by the existing heavy nuclei. 
The major isotope produced after oxygen burning is \Si.

Silicon burning follows the oxygen burning and is the final advanced burning stage that releases energy.
The temperature of silicon burning is about $3\,\times\,10^9\,\K$. In such high temperatures, energetic 
photons are able to disintegrate the heavy nuclei; this process is called photodisintegration. During
the silicon burning, part of the silicon is first photodisintegrated; the light isotopes are then recaptured 
by the silicon, and the resulting isotopes are photodisintegrated recursively. Such reactions build up 
a comprehensive reaction network and tend to reach a status called nuclear statistical equilibrium (NSE). 
The forward and backward reaction rates in NSE are almost equal. However, a perfect NSE occurs only at temperatures 
$>\,7\,\times\,10^9\,\K$. At the end, silicon burns into the iron group, including iron, cobalt, and nickel, and no more energy can 
be released from burning these isotopes. The major nuclear-burning reactions inside a star are listed in Table~\ref{ta_burn}. However, 
not every star goes through all of these burning processes; it depends on their initial masses.

  \begin{table}[ht]
   \tbl{Major burning processes: $T_{\rm min}$: the minimum temperature to ignite the burning \cite{prian2000}}
      {\begin{tabular}{@{}cccc@{}} \toprule
    Fuel   &	Reaction	 &  $T_{min}$[$10^6\,\K$] & yields   \\ \colrule
 H   &	$p-p$	 &  4 & He  \\
 H   &	CNO	 &  15 & He  \\
 He   &	$3\,\alpha$	 &  100 & C,O  \\
 C   &	C+C	 &  600 & O, Ne, Na, Mg  \\
 O   &	O+O	 &  1000 & Mg, S, P, Si  \\
 Si   &	NSE to iron group	 &  3000 & Co, Fe, Ni \\  \botrule
   \end{tabular}  \label{ta_burn} }
\end{table}

\begin{figure}[ht]
\includegraphics[width=\columnwidth]{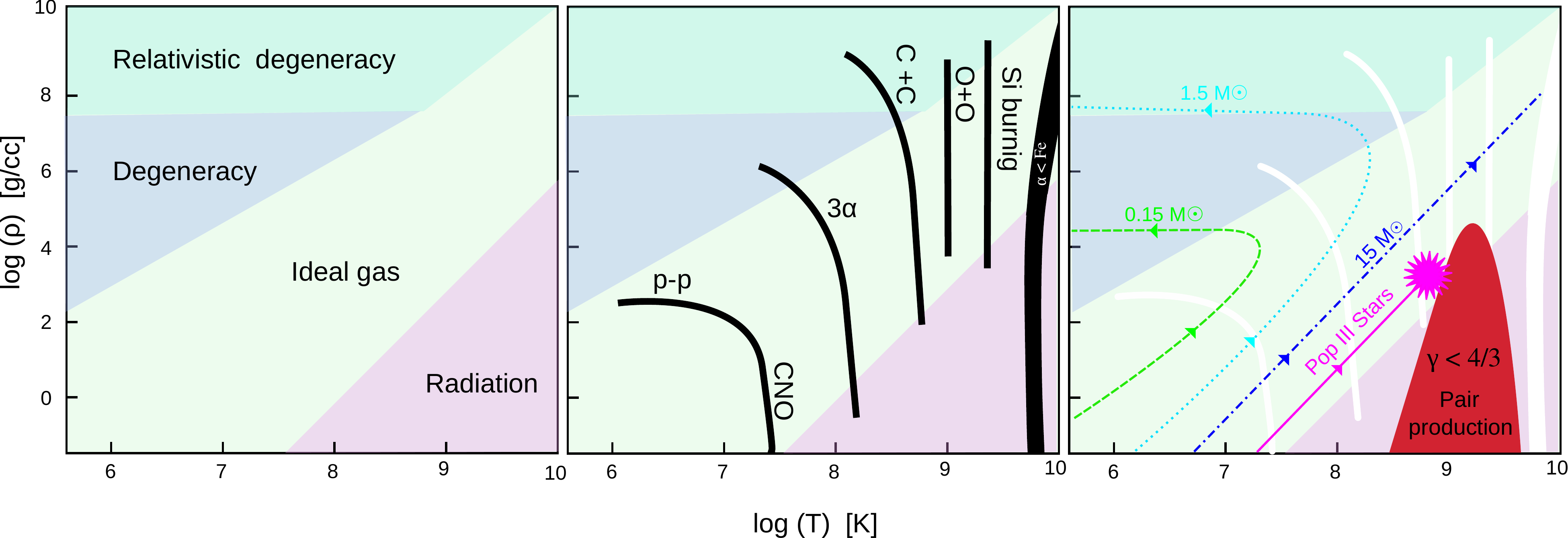}
\caption[]{Temperature-Density phase diagram based on [\refcite{prian2000}]: 
The x-axis and y-axis indicate the temperature and density, respectively. Colored patches show
the equation of state for matter (e.g., the radiation-dominated region (pink color) appears at a higher temperature with a lower density). 
In the middle panel, the black strips indicate the threshold for ignition of different burning phases in the phase diagram. In the right panel,  
the stellar evolution of stellar cores is shown in dashed lines. The red region shows the pair-instability region where the 
adiabatic index $\gamma_{\rm a}$ is below $\frac{4}{3}$.}  \label{trhoall_fig}
\end{figure}

Energetic photons may turn into electron-positron ($e^-/e^+$) pairs when they interact with the nucleus. 
The threshold energy of a photon for pair-production is $h\nu\,\sim\,2m_ec^2$, where 
$m_e$ is the rest mass of the electron, and $c$ is the speed of light. This energy
scale corresponds to a temperature of about $T\,\sim\,2m_ec^2/k_{\rm B}\sim10^{10}\,\K$.
At temperatures higher than  $10^{9}\,\K$, photons in the tail of the Planck distribution 
are energetic enough to create $e^+/e^{-}$ pairs. Pair production can lead to dynamical instabilities in the cores 
of stars because the pressure-supporting photons have become  
exhausted and turned into pairs. Pair-instabilities usually occur in very massive stars with
masses over $80\,\Msun$. If the temperature is sufficiently high, the stable iron 
group elements can also be photodisintegrated and break into $\alpha$ particles and 
neutrons. This process is called {\it iron photodisintegration}:
\begin{equation}
 \Fe\,+\,\gamma\,\rightarrow\,13\,\He\,+\,4n. 
\end{equation}
This reaction requires a photon energy over $100$~MeV. Helium becomes more abundant
than iron when the temperature rises over $7\times10^{9}\,\K$. Helium can be
disintegrated into neutrons and protons at even higher temperatures. In general,
the heavy nuclei are created at temperatures within $\sim10^6-5\times10^9\,\K$ through 
nuclear fusion and destroyed by energetic photons when the temperature is over $5\times 10^9\,\K$. 
Figure~\ref{trhoall_fig} summarizes the phase diagram of the stellar interior and burning and presents 
the schematic evolution tracks of stars of different masses. 
In the left panel, we show the density and the temperature phase diagram. 
When the relative lower density is subjected to high temperature, the equation for the state of gas can be described 
as ideal gas or radiation. For lower temperatures with a relatively higher density,
 quantum effects need to be considered for describing the equation of state. The gas can 
be degenerate or relativistic degenerate. In the middle panel, we show the different 
burning phases that occur in the phase diagram. The black strips show the approximate 
temperatures and densities when the burning occurs. We plot the evolution tracks
of central densities and temperatures of stars with different masses in
the right panel. The $0.15\,\Msun$ star may never reach the helium-burning stage before
its core becomes degenerate, and eventually it dies as a brown dwarf. The 
$1.5\,\Msun$ star, which is similar to our Sun, dies as a white dwarf after it finishes
the central helium burning. Once the star becomes more massive than $10\,\Msun$,
such as the $15\,\Msun$ star, it can go through all the burning stages we have mentioned, and it dies 
as an iron core$-$collapse supernovae (CCSNe). If the Pop~III stars were more massive than $80\,\Msun$,
they would encounter the pair-instabilities, which trigger a collapse of the stars, 
and they die as pair-instability supernovae.

We have mentioned several different fates of stars in the previous section. 
One common occurrence is that before the stars die, they encounter an instability 
that goes violent, the stars cannot restore it, and this leads to the catastrophic 
collapse of stars. It is relevant to provide an example of dynamical instability. 
Hydro$-$equilibrium means 
that the motion of fluid is too slow to be observed. To verify whether the state is a 
true equilibrium or not, we apply a perturbation to the equilibrium and 
evaluate the resulting response. The force balance inside a star is between the 
gravitational force and pressure gradient. In a simplified model, 
we consider a gas sphere of mass $M$, which is in a hydrodynamic equilibrium,
\begin{equation}
 \frac{dP}{dr}\,=\,-\rho\frac{Gm}{r^2},
\end{equation}
is equal to 
\begin{equation}
 \frac{dP}{dm}\,=\,-\frac{Gm}{4\pi\,r^4},
\end{equation}
in mass coordinate and its integration yields
\begin{equation}
 P\,=\,-\int^{M}_{m}\frac{Gm}{4\pi\,r^4} dm.
\label{press_eq}
 \end{equation}
Similar to [\refcite{prian2000}], we now perturb the system by compressing it by: 
\begin{equation}
\delta r\,=\,\alpha\,r,
\end{equation}
$\alpha\,\ll\,1$. Now the new density, $\tilde{\rho}$ and  radius, $\tilde{r}$ become
\begin{eqnarray*}
   \tilde{r} = r-\alpha\,r = r(1\,-\,\alpha),\\
  \tilde{\rho} = \frac{dm}{4\pi\,\tilde{r}^2 d\tilde{r}} \approx \rho(1\,+\,3\alpha).
\end{eqnarray*}
New pressure from hydrodynamics can be calculated by using the equation (\ref{press_eq})
\begin{equation}
\tilde{P}_h\,=\,\int^{M}_{m}\frac{Gm}{4\pi\,\tilde{r}^4} dm \,=\, \int^{M}_{m}\frac{Gm}{4\pi\,(1\,-\alpha\,)^4{r}^4} dm = (1\,+\,4\alpha)P.
\end{equation}
Assuming the contraction is adiabatic, the gas pressure can be expressed as 
\begin{equation}
\tilde{P}_{\rm gas}\,=\,K_{\rm a}\tilde{\rho}^{\,\gamma_{\rm a}}\,=\,K_{\rm a}[\rho(1\,+\,3\alpha)]^{\gamma_{\rm a}}\,=\, (1\,+\,3\gamma_{\rm a}\alpha)P,
\end{equation}
where $K_{\rm a}$ is a constant. The contraction of the gas sphere can be restored when 
\begin{equation}
\tilde{P}_{\rm gas}\,>\,\tilde{P}_h \quad \longrightarrow \quad (1\,+\,3\gamma_{\rm a}\alpha)P \,>\, (1\,+\,4\alpha)P.
\end{equation}
Therefore, the condition for dynamical stability is 
\begin{equation}
\gamma_{\rm a}\,>\,4/3, 
\end{equation}
which can be further extended to a global stability, 
\begin{equation}
\int(\gamma_{\rm a}-4/3)\frac{P}{\rho}\,dm\,>\,4/3,
\end{equation}
which implies that the star can be stable if $\gamma_{\rm a}\,>\,4/3$ occurs in the region
where $P/\rho$ is dominated, e.g., the core of the star; even the outer envelope
may have $\gamma_{\rm a}\,<\,4/3$. 

\subsection{Supernovae Explosions}
The fate of a massive star is determined by its initial mass, 
composition, and history of mass loss. The mechanics of mass loss is  poorly understood. 
The explosion mechanism and remnant properties are thought to be determined by the 
mass of the helium core at the time before the star dies. [\refcite{kudri2002}] suggests that 
the mass loss rate of a star follows $\dot{m}\,\propto\,Z^{0.5}$, where $Z$ is the metallicity 
of a star relative to the solar metallicity, $\Zsun$. Since the 
Pop~III stars have zero metallicity, it would favor the notion that Pop~III stars retain most 
of their masses before they die. The Pop~III stars with initial masses of $10 - 80\,\Msun$ 
eventually forge an iron core with masses similar to those of our Sun\cite{kippen1990}{}. 
Once the mass of the iron core is larger 
than its Chandrasekhar mass\cite{chand1942}{}, the degenerate pressure of electrons can no longer support 
the gravity from the mass of the core itself; these conditions trigger the dramatic implosion of the core and compress the core 
to nucleon densities of
about $10^{14}\,\gcc$. Most of the gravitational energy is released  in the form of energetic neutrinos, which eventually power 
the CCSNe. The core of the star then collapses into a neutron star or a black hole, depending on the mass of the progenitor star 
\cite{woosley1986,woosley2002,woosley2005}{}. The neutrino-driven explosion mechanism for CCSNe is still poorly understood 
because it is complicated by issues of micro-physics, multi-scale, and multi-dimension [\refcite{burrows1995,janka1996,mezz1998,murphy2008,nordhaus2010}]. It is predicted that only about $1\%$ of the energy from neutrinos goes into the SN ejecta, 
which shines as brightly as the galaxy for a few weeks before fading away.
In recent decades, theorists and observers have been
fascinated by many different aspects of CCSNe, such as the explosion mechanisms, nucleosynthesis, compact remnant, etc. The photons from CCSNe carry information about their progenitor stars as well as their host galaxies, which 
makes CCSNe a powerful tool for studying the Universe.

\begin{figure}[ht]
\begin{center}
\includegraphics[width = \columnwidth]{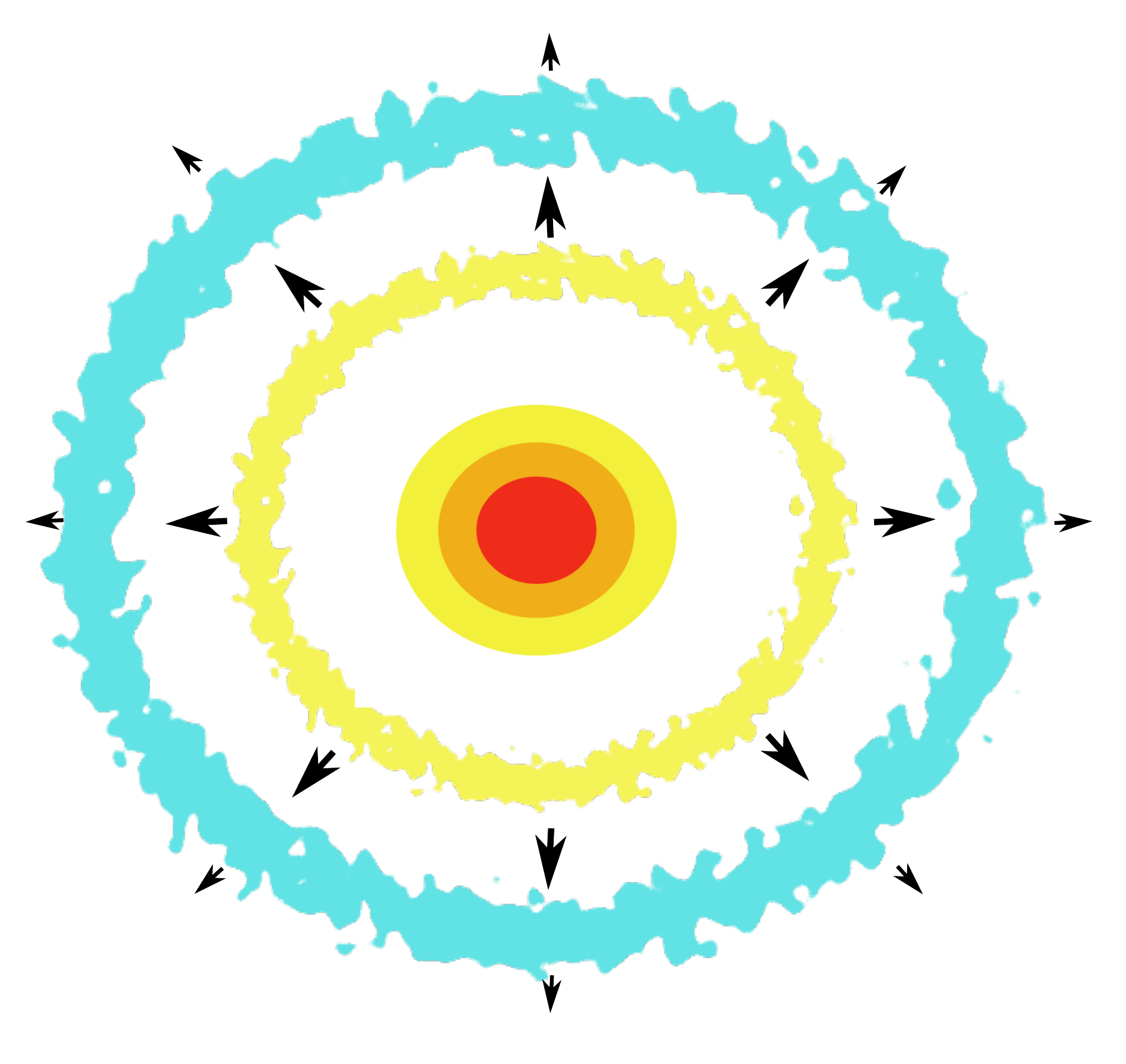}
\caption[]{ Illustration of PPSNe: During the core collapsed trigger by pair-instability, 
the energy released from central oxygen burning is not sufficient to disperse the star but can easily eject masses from its envelope. A few outbursts of mass 
can occur before the star dies as a CCSN. The latter outbursts are more energetic than the earlier ones, that leads to the collision of ejecta. \label{ppsn_cartoon}}
\end{center}
\end{figure}

If Pop~III stars are more massive than $80\,\Msun$, after the central carbon burning, 
their cores encounter the $e^-/e^+$ pair production instabilities, in which large amounts of 
pressure-supporting photons are turned into $e^-/e^+$ pairs, leading to dynamical instability 
of the core. The central temperatures start to oscillate. If the stars are more massive than 
$100\,\Msun$, the oscillation of temperatures becomes very violent. Several strong shocks may be 
sent out from the core before the stars die as CCSNe \cite{woosley2007}{}. Those shocks are 
inadequate to blow up the entire star, but they are strong enough to eject several solar masses 
from the stellar envelope, as is illustrated in Figure~\ref{ppsn_cartoon}.  The collisions of ejected mass 
may power extremely luminous optical transients, the which are called pulsational pair-instability supernovae (PPSNe). 

\begin{figure}[ht]
\begin{center}
\includegraphics[width=\columnwidth]{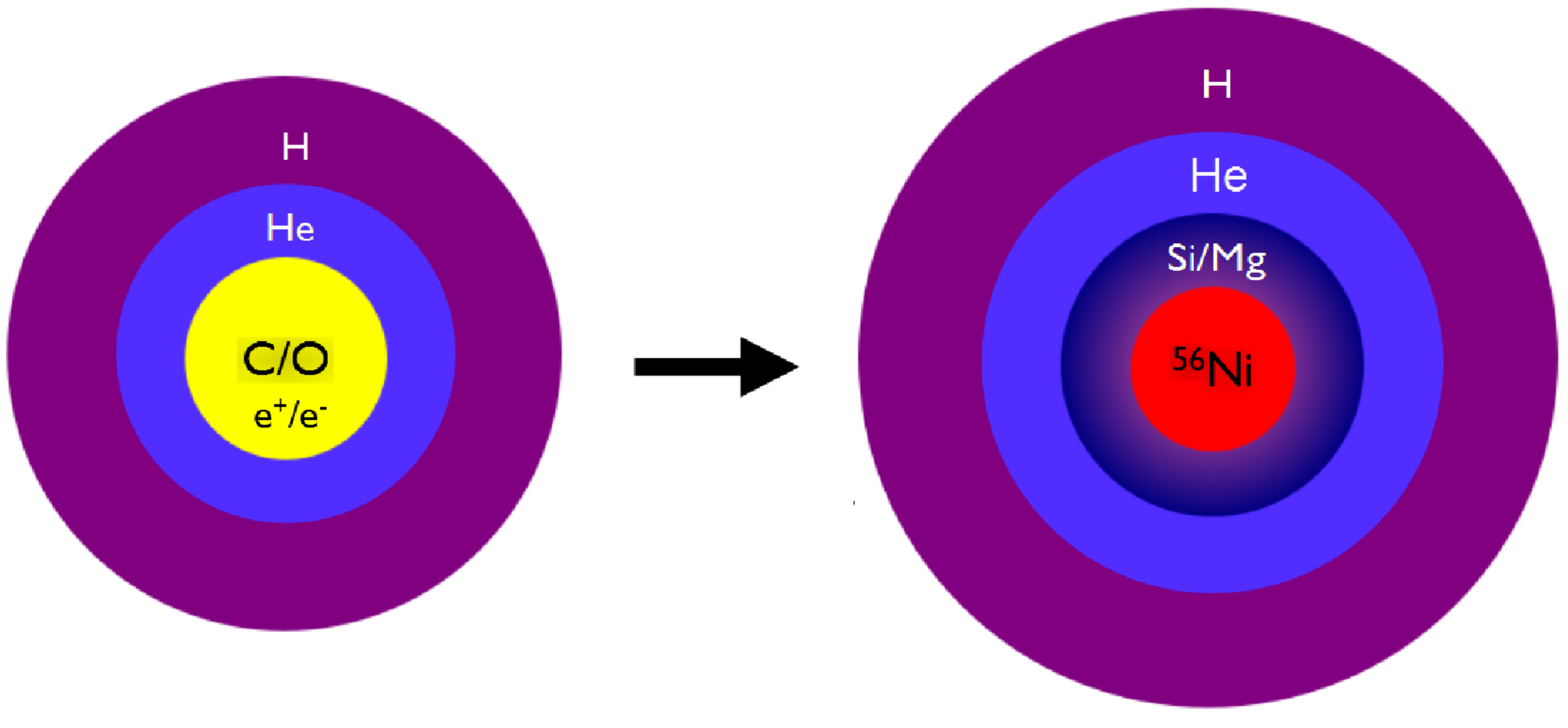}
\caption[]{ Illustration of PSNe: After central helium burning, pressure-supporting photons 
of core are converted into $e^-/e^+$; the core becomes dynamically unstable, resulting 
in an implosion that ignites the oxygen and silicon burning explosively. The energy released from 
burning  totally blows up the star, and some amounts of \Ni{} are synthesized. (Image credit: Dan Kasen)\label{psn_cartoon}}
\end{center}
\end{figure}

Once the stars are over $150\,\Msun$ but less than $260\,\Msun$, instabilities are so violent they trigger a runaway collapse 
and eventually ignite the explosive oxygen and silicon burning, resulting in an energetic explosion and completely disrupting the star, 
as shown in Figure~\ref{psn_cartoon}. This thermonuclear explosion is called a pair-instability supernova. A PSN can produce an explosion energy up to $\Ep{53}\,\erg$, about 100 times more energetic than the Type Ia SNe. Because of explosive silicon burning, a large amount of radioactive \Ni{} is synthesized. Such an energetic explosion
makes them very bright, and they can be visible at large distances, so they may function as good tools for probing the early Universe. For the 
yields of PSNe, isotopes heavier than the iron group are completely absent because of a lack of neutron capture processes 
(r- and s-process). 

What happens to even more massive stars? Previous models suggest that non-rotating stars with initial masses over $260\,\Msun$ 
eventually die as BH without SN explosions.   It is generally believed that the explosive burning is insufficient 
to revert the implosion  because the SN shock is dissipated by the photo-disintegration of the heavy nuclei; thus these stars 
eventually die as BHs without SN explosions. However, [\refcite{chen_phd}] reported an unusual explosion of a super massive star with a mass 
about $55,000\,\Msun$. This unexpected explosion may have caused the post-Newtonian correction in the gravity. We summarize the fate of massive Pop~III stars in Table~\ref{ta2} based on [\refcite{woosley2002,heger2010}].

In this review, we focus on the  (pulsational) pair-instability supernovae and 
possible explosions among the extremely massive stars.  Most current theoretical 
models of these are based on one-dimensional calculations.  
Only  very recently have results from multi-D models become available. 
In the initial stages of a supernova, however, spherical symmetry may be broken by 
fluid instabilities generated by burning that cannot be captured in 1D.  The mixing 
due to fluid instabilities may be able to affect the observational signatures of these 
SNe. We will discuss some of the latest multidimensional models of these Pop~III 
SNe. 

 \begin{table}[ht]
   \tbl{Death of Massive Stars}
   {\begin{tabular}{@{}lll@{}} \toprule
   \MS [\Msun]		 & He core [\Msun] & Supernova Mechanism   \\ \colrule
       $10\hphantom{0} \quad\leftrightarrow\quad 85$ & $2\hphantom{00}\quad\leftrightarrow\quad32$  & 	CCSNe\\
       $80\hphantom{0} \quad\leftrightarrow\quad 150$ & $35\hphantom{0}\quad\leftrightarrow\quad60$  & 	PPSNe\\
       $150 \quad \leftrightarrow\quad 260$ & $60\hphantom{0}\quad\leftrightarrow\quad133$ & 	PSNe\\
        $\hphantom{000} \quad \ge \quad \, 260$ &    $\hphantom{000} \quad \ge \quad\,133$ & 	BHs (?) \\ \botrule
      \end{tabular} \label{ta2}}
   \end{table}

\section{Supernova Explosions with \CASTRO}
\label{sn_castro}
Multidimensional SN simulations are usually computationally expensive and 
technically difficult, requiring a robust code and powerful supercomputers 
to realize. In this section, we introduce our modified version of \CASTRO{} 
which is designed for such problems. \CASTRO{} \cite{ann2010,zhang2011}{} is  
a massively parallel, multidimensional Eulerian, adaptive mesh refinement (AMR),
hydrodynamics code for astrophysical applications. The code was originally developed at 
the Lawrence Berkeley Lab, and it is designed to run effectively on supercomputers of 
10,000+ CPUs. \CASTRO{} provides a powerful platform for simulating hydrodynamics and gravity
for astrophysical gas dynamics. However, it still requires other physics to properly 
model supernova explosions. We review some of the key physics and associated numerical 
algorithms.

The structure of this section is as follows:  we
first describe features of \CASTRO{} in \S~\ref{castro_sec}, then 
introduce the nuclear reaction network in \S~\ref{burning_sec}. 
The algorithms for the 1D-to-MultiD Mapping are 
presented in \S~\ref{mapping_sec}. We discuss post-Newtonian gravity in \S~\ref{gr_sec} and 
an approach for resolving the large dynamic scale of simulations in 
\S~\ref{resolve_sec}. At the end, we present the scaling performance of 
\CASTRO{} in \S~\ref{castro_mpi_sec} and introduce \VISIT{}, the tool for visualizing
\CASTRO{} output, in \S~\ref{visit_sec}.

\subsection{\CASTRO}
\label{castro_sec}
\CASTRO{} is a hydro code for solving compressible hydrodynamic equations of multi-components including self-gravity and 
a general equation of state (EOS). The Eulerian grid of \CASTRO{} uses adaptive mesh refinement (AMR), which constructs 
rectangular refinement grids hierarchically. Different coordinate systems are available in \CASTRO{}, including 
spherical (1D), cylindrical (2D), and cartesian (3D). The flexible modules of \CASTRO{} make it easy for users to implement 
new physics associated with their simulations. 

In \CASTRO, the hydrodynamics are evolved by solving the conservation equations of mass, momentum, and energy [\refcite{ann2010}] :
\begin{eqnarray}
\frac{\partial \rho}{\partial t} &=& - \nabla \cdot (\rho \ub), \\
\frac{\partial (\rho \ub)}{\partial t} &=& - \nabla \cdot (\rho \ub \ub) - \nabla p + \rho \gb, \\
\frac{\partial (\rho E)}{\partial t} &=& - \nabla \cdot (\rho \ub E + p \ub) 
+ \rho \dot{\epsilon}_{\rm nuc} 
+ \rho \ub \cdot \gb. \lEq{e2},
\end{eqnarray}
where $\rho$, $\ub$, $e$, and $E$ are the mass density, velocity vector, internal energy per unit mass, 
and total energy per unit mass $E = e + \ub \cdot \ub / 2$, respectively. The pressure, $p$, is calculated
from the equation of state (EOS), $\gb$ is the gravity, and $\dot{\epsilon}_{\rm nuc}$ is the energy generation 
rate per unit volume. \CASTRO{} also evolves the reacting flow by considering the 
advection equations of the mass abundances of isotopes, $X_i$ : 
\begin{equation}
\frac{\partial (\rho X_i)}{\partial t} = - \nabla \cdot (\rho \ub X_i) + \rho \dot\omega_i,
\lEq{sp1}
\end{equation}
where $\dot\omega_i$ is the production rate for the $i$-th isotope having the form:
\begin{equation}
 \omegadot_i(\rho,X_i,T)= \frac{dX_i}{dt},
\end{equation}
is given from the nuclear reaction network that we shall describe later. 
Since masses are conservative quantities, the mass fractions are subject 
to the constraint that $\sum\limits_{i} X_i = 1$.  
\CASTRO{} can support any general reaction network that takes as inputs the density, temperature, 
and mass fractions of isotopes, and it returns updated mass fractions and the energy  generation rates.  
The input temperature is computed from the EOS before each call to the reaction network.  At the end of 
the burning step, the results of burning provide the rates of energy generation/loss and abundance change 
to update \Eq{e2} and \Eq{sp1}. \CASTRO{} also provides passively advected quantities; $A_j$, e.g., angular 
momentum, which is used for rotation models,
\begin{equation}
\frac{\partial (\rho A_j)}{\partial t} = - \nabla \cdot (\rho \ub A_j).
\end{equation}
\CASTRO{} uses a sophisticated EOS  for stellar matter: the Helmholtz
\cite{timmes2000}{}, which considers the (non)degenerate and (non)relativistic 
electrons, electron-positron pair production, as well as ideal gas with radiation. 
The Helmholtz EOS is a tabular EOS that reads in $\rho$, $T$, and $X_i$ of gas and 
yields its derived thermodynamics quantities. \CASTRO{} offers different types
of calculation for gravity, including Constant, Poisson, and Monopole. 
At the early stage of a supernova explosion, spherical symmetry is still a good 
approximation for the mass distribution of gas. Such an approximation creates a 
great advantage in calculating the gravity by saving a lot of computational time, 
so the monopole-type gravity is usually used in the simulations.
In multidimensional \CASTRO{} simulations, we first calculate a 1D radial average profile of density. 
We then compute the 1D profile of $\gb$ and use it to calculate the gravity of the multidimensional 
grid cells. 

\begin{figure}[ht]
\begin{center} 
\subfigure[AMR cartoon]{\label{amr_a_fig}\includegraphics[width=0.45\textwidth]{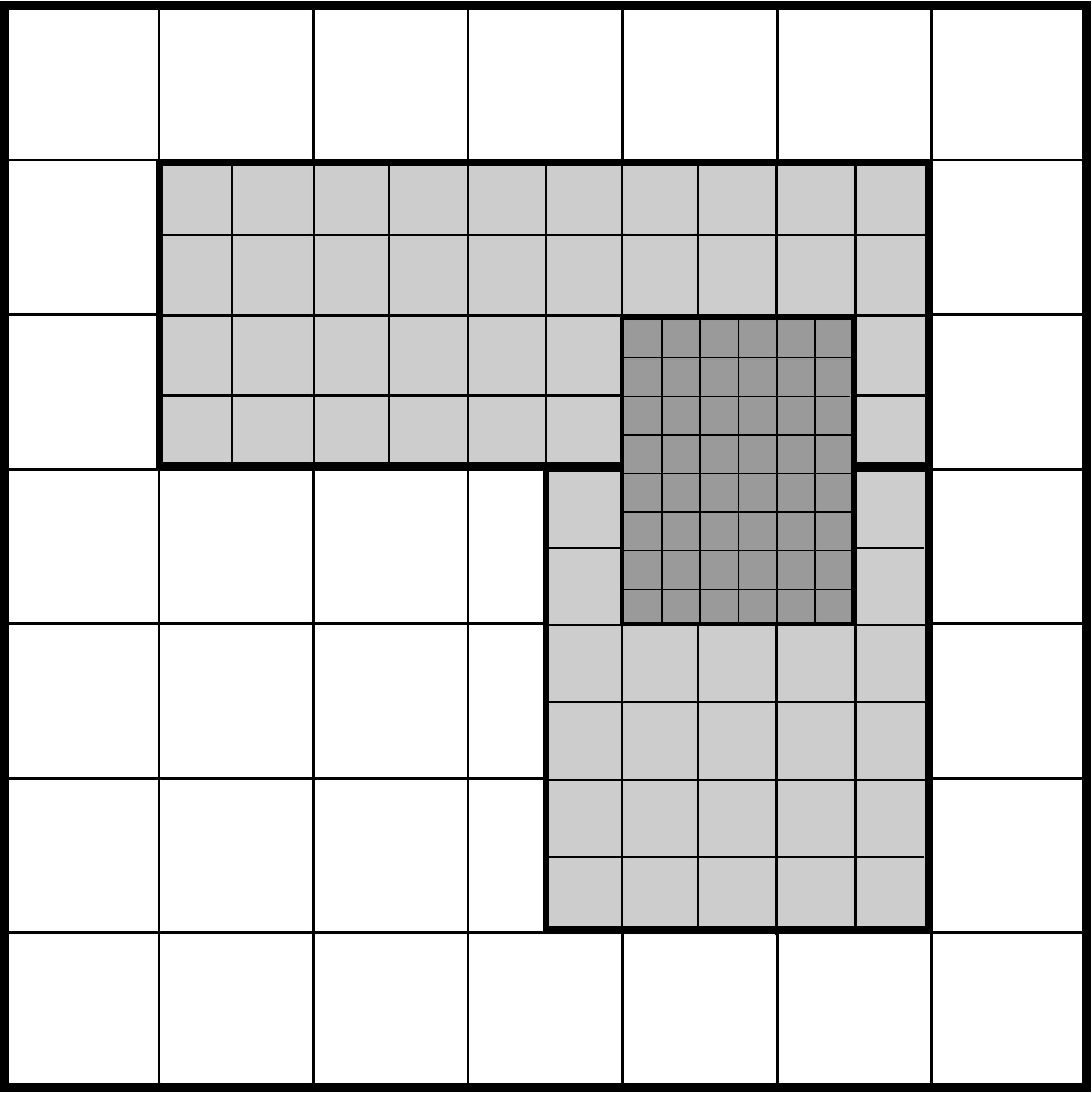}}                
\subfigure[3D AMR]{\label{amr_b_fig} \includegraphics[width=0.45\textwidth]{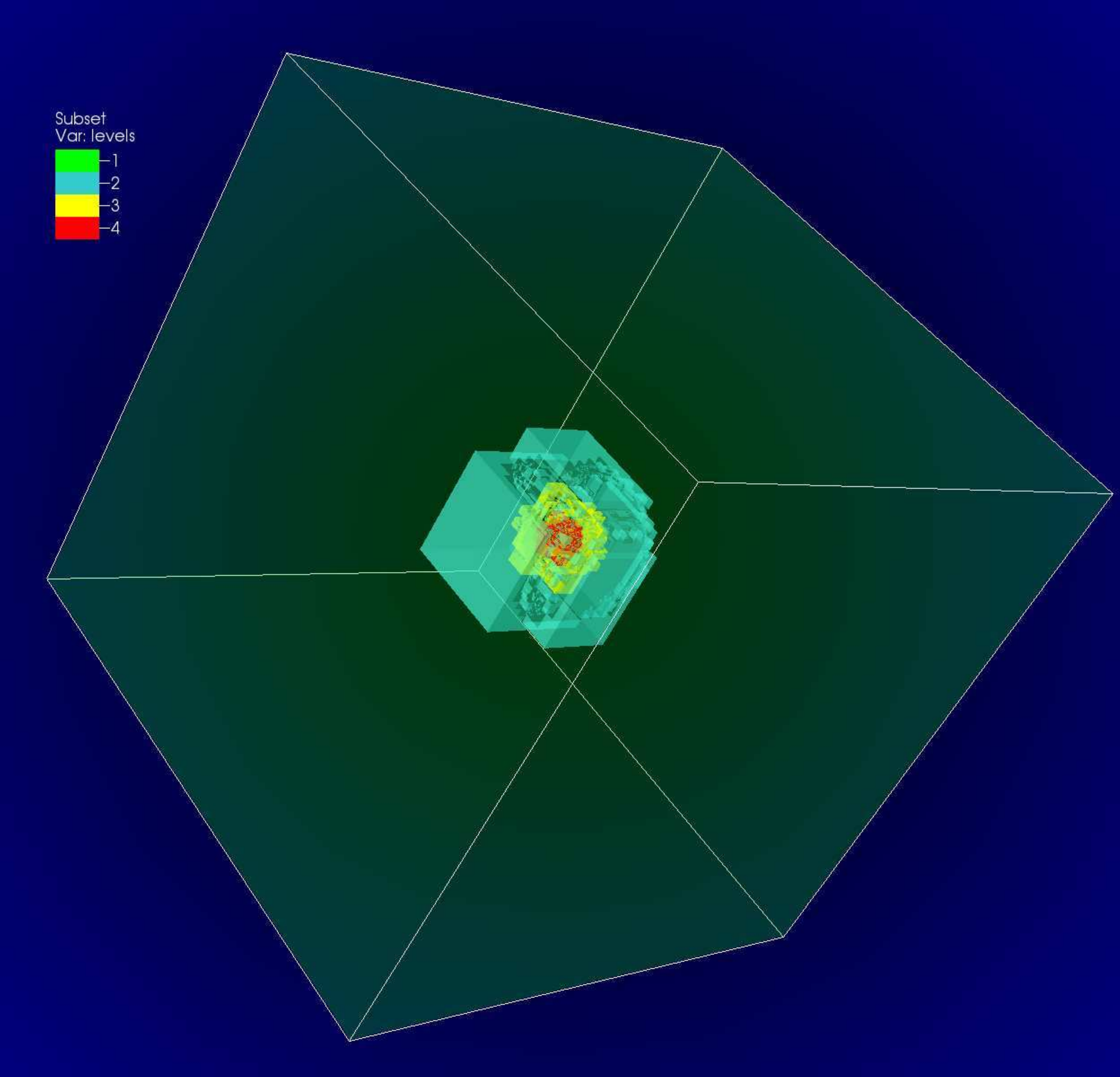}}   
\caption{(a) Schematic diagram of AMR: When the refinement criteria are met, new finer grids are automatically generated to replace previous coarse grids. Two levels of refinement are shown in 
light-gray and dark-gray on the top of coarse grids (white). (b) 3 D nested grids: The hierarchical 
grids are constructed from the center and the colors represent three different levels of refinement. It is 
very useful for importing a 3D SN onto such layout of grids, where the core of SN can have the highest
spacial resolutions all the time.}
\end{center}
\end{figure}

The AMR in \CASTRO{} refines the simulation domain in both space and time. 
Finer grids automatically replace coarse grids during the grid-refining process 
until the solution satisfies the AMR criteria, which are specified by users. 
These criteria can be the gradients of densities, velocities, or other physical 
quantities in the adjacent grids. The grid generation procedures automatically create 
or remove finer rectangular zones based on the refinement criteria. 
The AMR technique of \CASTRO{} allows us to address our supernova simulation, which deals
with a large dynamic scale. Simulating the mixing of supernova ejecta requires 
catching the features of fluid instabilities early on. These instabilities occur 
at much smaller scales compared with the overall simulation box. The uniform grid
approach requires numerous zones and becomes very computationally expensive. Instead,
AMR focuses on resolving the scale of interests and makes our simulations run more
efficiently. In Figure~\ref{amr_a_fig}, we show the layout of two levels of a 
factor of two refinement. The refined grids are constructed hierarchically 
in the form of rectangles. The choice of refinement criteria allows us to
resolve the structure we are most interested in. The most violent burning
and physical process occurs at the center of the star, so we usually apply
hierarchically-configured zones at the center of simulated domain, as shown in Figure~\ref{amr_b_fig}. 
These pre-refined zones are fixed and do not change with AMR criteria.

\begin{figure}[ht]
\begin{center}
\includegraphics[width=\columnwidth]{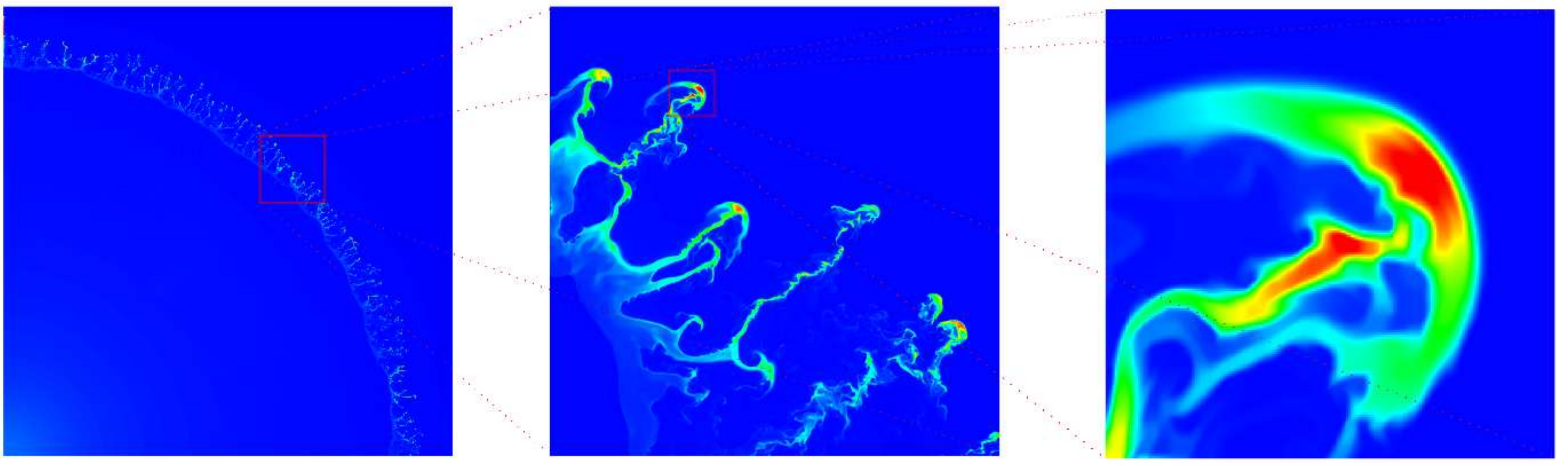}
\caption[Power of AMR]{Ultra-high resolution with AMR: This is a snapshot of the density 
from 2D \CASTRO{} simulations\cite{chen_phd} at the time when the fluid instabilities emerge.
The images from left to right show the magnification of the instabilities. With AMR, the detailed 
structures of fluid instabilities are fully resolved. \label{amrp}}
\end{center}
\end{figure}

Figure~\ref{amrp} shows the power of AMR in the simulations. This is a snapshot taken from 
our 2D supernova simulation at the time when the fluid instabilities emerge. These fluid 
instabilities are caused by Rayleigh--Taylor (RT) instability and are the main drivers of the mixing of SN ejecta.
The finest grids of AMR can resolve the detailed structure of fluid instabilities at minimal 
computational expense. In our simulations, AMR criteria are based on density gradient, velocity gradient, 
and pressure gradient.

\subsection{Nuclear Reaction Networks}
\label{burning_sec} 
Modeling thermonuclear supernovae requires calculating the energy generation 
rate from nuclear burning, which occurs over a large range of temperatures, 
densities, and compositions.  We have implemented the APPROX 7, 13, 19$-$ isotope 
reaction networks\cite{kepler,timmes1999} into \CASTRO.  Here, 
we introduce the 19 isotopes reaction network, which is the most comprehensive 
network afforded for multidimensional simulations.  This network includes 
19 isotopes:  \Hy, \Hea, \Heb, \Cx, \Nx, \Ox, \Ne, \Mg, \Si, \Sx, \Ar, \Ca, \Ti, 
\Cr, \iFe, \jFe, \Ni, protons (from photo-disintegration), and neutrons.  
The 19$-$isotope network considers nuclear burning of alpha-chain reactions, 
heavy-ion reactions, hot CNO cycles, photo-disintegration of heavy elements, and 
neutrino energy loss.  It is capable of 
efficiently calculating accurate energy generation rates for nuclear processes 
ranging from hydrogen to silicon burning.

The nuclear reaction networks are solved by means of integrating a system of ordinary differential 
equations.  Because the reaction rates for most of the burning are extremely sensitive to temperatures
to $\propto T^{15-40}$, it results in stiffness of the system of equations, which are usually solved by 
an implicit time integration scheme. We first consider the gas containing $m$ isotopes with a density 
$\rho$ and temperature $T$.  The molar abundance of the $i$-th isotope is
\begin{equation}
Y_i = \frac{X_i} {A_i} = \frac{\rho_i}{\rho A_i} =\frac{n_i}{\rho N_A},
\end{equation}
where  $A_i$ is mass number, $X_i$ is mass fraction, $\rho_i$ is mass density, and 
$N_A$  is the Avogadro's number. In Lagrangian coordinates, the continuity equation
of the isotope has the form\cite{timmes1999}
\begin{equation}
\frac{dY_i}{dt}+\nabla\cdot(Y_i{\bf V}_i) = \dot{R_i},
\lEq{cont}
\end{equation}
where
\begin{equation}
\dot{R_i} = \sum_{j,k} Y_l Y_k\lambda_{kj}(l) -Y_i Y_j\lambda_{jk}(i),
\lEq{rate}
\end{equation} 
where $\dot{R_i}$ is the total reaction rate due to all binary reactions of the form $i(j,k)l$. 
$\lambda_{jk}$  and $\lambda_{kj}$  are the  forward and reverse nuclear reaction rates, 
which usually have a strong temperature dependence. 
${\bf V}_i$  are mass diffusion velocities due to pressure, temperature, and abundance gradients.
The value of ${\bf V}_i$ is often small compared with other transport processes, 
so we can assume ${\bf V}_i = 0$, which allows us to decouple the reaction network 
from the hydrodynamics by using operator splitting.  \Eq{cont} now becomes
\begin{equation}
\frac{d { Y_i}}{dt} = \dot{R_i}.
\lEq{cont2}
\end{equation} 
This set of ordinary differential equations may be written in the more compact and standard form\cite{timmes1999}
\begin{equation}
\frac{d{\bf y}}{dt} = {\bf f(y)};
\lEq{cont3}
\end{equation}
its implicit differentiation gives 
\begin{equation}
{\bf y}_{n+1} = {\bf y}_{n} +h{\bf f(y}_{n+1}),
\lEq{s1}
\end{equation}
where $h$ is a small time step.  We linearize \Eq{s1} by using Newton's method,
\begin{equation}
{\bf y}_{n+1} = {\bf y}_{n} +h\left[{\bf f(y}_{n})+{\frac{\bf \partial{f}}{\partial{\bf y}}}\bigg\vert_{\bf{y}_n}\cdot({\bf y}_{n+1} - {\bf y}_{n})\right].
\lEq{s2}
\end{equation}
The rearranged \Eq{s2} yields 
\begin{equation}
{\bf y}_{n+1} = {\bf y}_{n} +h \left[{\bf 1}-h \frac{\partial{\bf f}}{\partial{\bf y}}\right]^{-1}\cdot {\bf f(y}_{n}).
\lEq{s3}
\end{equation}
By defining ${\Delta} = {\bf y}_{n+1} -{\bf y}_{n} $, $\tilde{\bf A} = \frac{\bf 1}{h}- \frac{\partial{\bf f}}{\partial{\bf y}}$,
$ {\bf b} = {\bf f(y}_{n})$, \Eq{s3} now is equivalent  to a simple matrix equation 
\begin{equation}
\tilde{\bf A} \cdot{\Delta} = {\bf b}.
\lEq{s4}
\end{equation}
If $h$ is small enough, only one iteration of Newton's method may be accurate 
enough to solve  \Eq{cont3} using \Eq{s3}.  However, this method 
provides no estimate of how accurate the integration step is.  We also do not know 
whether the time step is accurate enough.  The Jacobian matrices 
$\tilde{\bf J} = \frac{\partial{\bf f}}{\partial{\bf y}}$ from nuclear reaction networks are 
neither positive-definite nor symmetric, and the magnitudes of the matrix elements are functions 
$X(t)$, $T(t)$, and $\rho(t)$. More importantly, the nuclear reaction rates are extremely sensitive 
to temperature, and $X$ of different isotopes can differ by many orders of magnitude. The coefficients in 
\Eq{cont2} can vary significantly and cause nuclear reaction network equations 
to become {\bf stiff}. 

The integration method for our network is based on a variable-order Bader--Deuflhard method\cite{press2007}{}. [\refcite{bader1983}] found a semi-implicit discretization for stiff equation problems and obtained an implicit form of the midpoint rule,
\begin{equation}
{\bf y}_{n+1} - {\bf y}_{n-1} =2h{\bf f}(\frac{{\bf y}_{n+1} + {\bf y}_{n-1}}{2}).
\lEq{d1}
\end{equation}
We linearize the right-hand side about ${\bf f(y}_n)$ and obtain the semi-implicit midpoint rule
\begin{equation}
\left[{\bf 1}-h\frac{\partial{\bf f}}{\partial{\bf y}}\right]\cdot{\bf y}_{n+1} = \left[{\bf 1}+ h\frac{\partial{\bf f}}{\partial{\bf y}}\right]\cdot{\bf y}_{n-1} 
+2h\left[{\bf f(y}_{n})- \frac{\partial{\bf f}}{\partial{\bf y}} \cdot \bf{y_n}\right].
\lEq{d2} 
\end{equation}
 Now the reaction network expressed in \Eq{cont3} is advanced over a large time step, $H=mh$ for ${\bf y}_{n}$ to 
${\bf y}_{n+1}$, where $m$ is an integer. It is convenient to rewrite equations in terms of 
$\Delta_k \equiv {\bf y}_{k+1} - {\bf y}_{k}$. We use it with the first step from \Eq{s3} and start by calculating\cite{press2007}
\begin{equation}
\begin{array}{l}
{\bf y}_{1} =  {\bf y}_{0} + \Delta_0, \\ \\
\Delta_0 = \left[{\bf 1}-h\frac{\partial{\bf f}}{\partial{\bf y}}\right]^{-1}\cdot h{\bf f(y}_{0}).
\end{array}
\lEq{d3} 
\end{equation}
Then for $k = 1,...,m -1$, set
\begin{equation}
\begin{array}{l}
{\bf y}_{k+1} =  {\bf y}_{k} + \Delta_k, \\ \\
\Delta_k = \Delta_{k-1}+2 \left[{\bf 1}-h\frac{\partial{\bf f}}{\partial{\bf y}}\right]^{-1}\cdot [h{\bf f(y}_{k})-\Delta_{k-1}].
\end{array}
\lEq{d3} 
\end{equation}
Finally, we calculate
\begin{equation}
\begin{array}{l}
{\bf y}_{n+1} =  {\bf y}_{m} + \Delta_m, \\ \\
\Delta_m = \left[{\bf 1}-h\frac{\partial{\bf f}}{\partial{\bf y}}\right]^{-1}\cdot [h{\bf f(y}_{m})-\Delta_{m-1}].
\end{array}
\lEq{d4} 
\end{equation}
This sequence may be executed a maximum of $7$ times, which yields a 15th-order method.  
The exact number of times the staged sequence is executed depends on the accuracy 
requirements. The accuracy of an integration step is calculated by comparing the 
solutions derived from different orders. The linear algebra package \GIFT{}\cite{muller1998} 
and the sparse storage package \MA{}\cite{duff1986} are used 
to execute the semi-implicit time integration methods described above. After solving 
the network equations, the average nuclear energy generated rate is calculated,
\begin{equation}
\dot{\epsilon}_{\rm nuc} = \sum_i \frac{\Delta Y_i}{\Delta t}B_iN_A-\dot{\epsilon}_{\nu}, 
\lEq{e1} 
\end{equation}
where $B_i$ is the nuclear binding energy of the $i$-th isotope, and $\dot{\epsilon}_{\nu}$ 
is the energy loss rate due to neutrinos\cite{itoh1996}{}. 

\subsection{Mapping}
\label{mapping_sec}
\begin{figure}[ht]
\begin{center}
\includegraphics[width=\columnwidth]{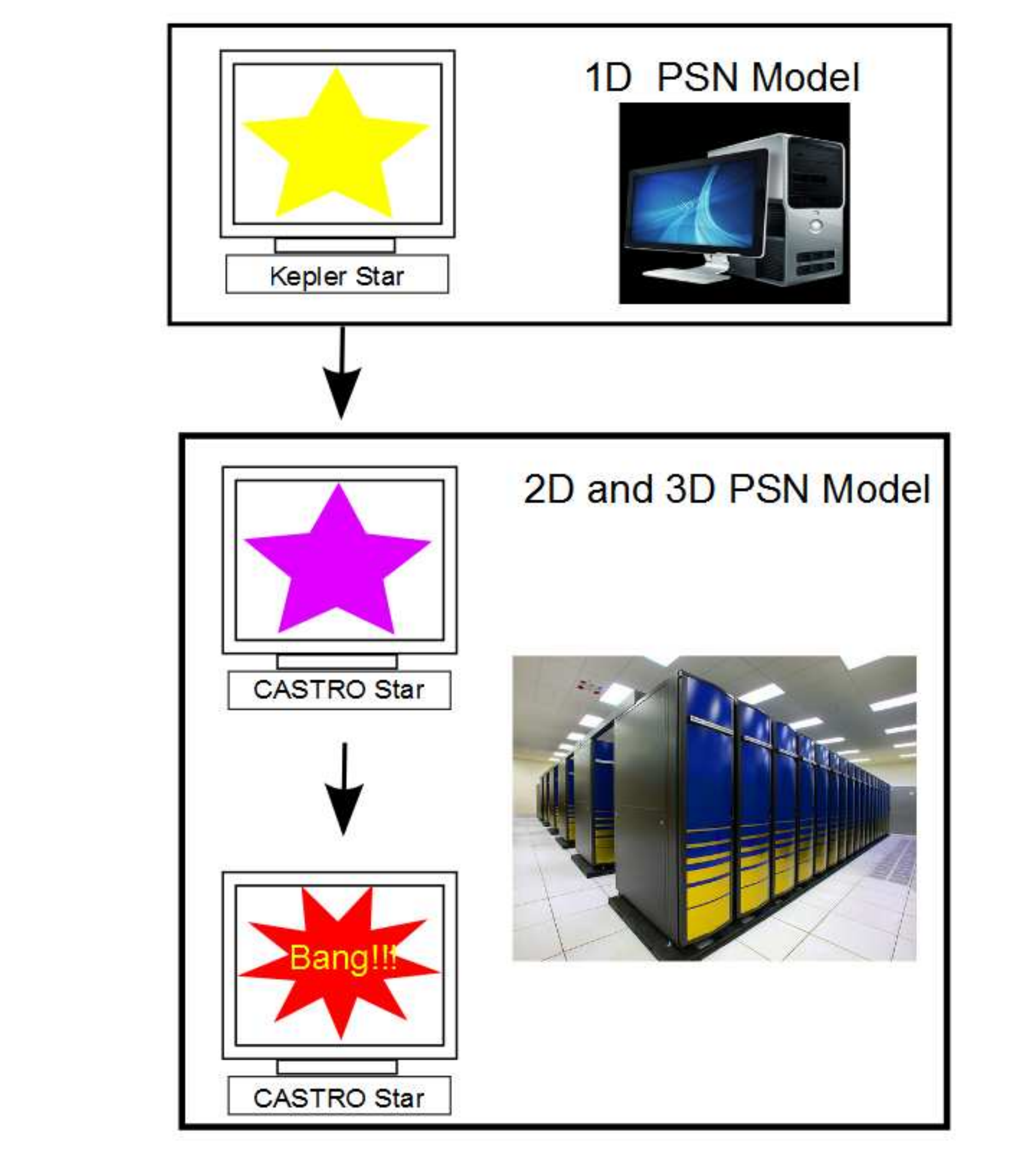}
\caption[1D to multi-D]{Procedure of multi-D SN simulations: The 1D stellar models can be  
generated by \KEPLER{} on personal computers. The resulting 1D supernova-progenitor 
models are mapped onto multidimensional grids of \CASTRO{}. Due to the intensive computation 
of multidimensional models, simulations require to run on supercomputers such as Franklin. Simulations usually stop after the star explodes.\label{models_fig}}
\end{center}
\end{figure}

Computing fully self-consistent 3D stellar evolution models, from their formation to collapse
for the explosion setup is unavailable in terms of current supercomputer capability.  One alternative approach 
is to first evolve the main sequence star in 1D stellar evolution codes such as \KEPLER{} \cite{kepler} 
or \MESA{} \cite{mesa}{}.  Once the star reaches the pre-supernova phase, its 1D profiles can then be 
mapped into multidimensional hydro codes such as \CASTRO{} or \FLASH \cite{flash} and 
continue to be evolved until the star explodes, as shown in Figure~\ref{models_fig}.

Differences between codes in dimensionality and coordinate mesh can lead to numerical issues 
such as violation of conservation of mass and energy when data are mapped from one code 
to another.  A first, simple approach could be to initialize multidimensional grids by linear 
interpolation from corresponding mesh points on the 1D profiles.  However, linear interpolation 
becomes invalid when the new grid fails to resolve critical features in the original profile, such as 
the inner core of a star.  This is especially true when porting profiles from 1D Lagrangian codes, 
which can easily resolve very small spatial features in mass coordinate, to a fixed or adaptive 
Eulerian grid. In addition to conservation laws, some physical processes, such as nuclear burning, are 
very sensitive to temperature, so errors in mapping can lead to very different outcomes for 
the simulations, including altering the nucleosynthesis and energetics of SNe. [\refcite{zingale2002}]  has 
examined mapping 1D profiles to 2D or 3D meshes under a hydro equilibrium status and 
[\refcite{mapping_chen}] has developed a new mapping scheme to conservatively map the 1D 
initial conditions onto multidimensional zones.

\begin{figure}[ht]
\begin{center}
\includegraphics[width=\columnwidth]{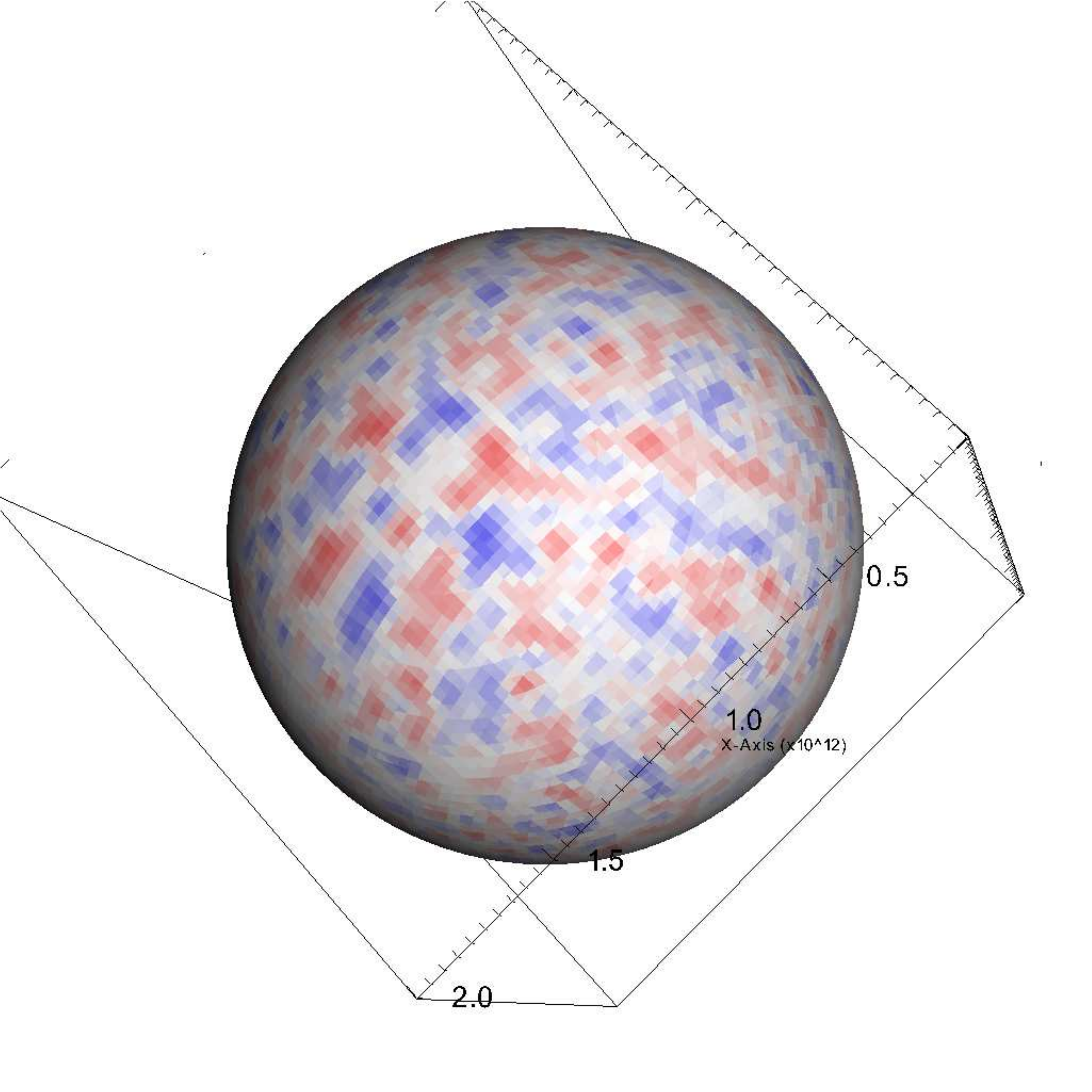}
\caption[3D perturbations]{ Turbulent perturbations: The sphere represents the convective core of  the star. Red and blue
colors show the positive and radial perturbed velocities, respectively.  \label{perb2}}
\end{center}
\end{figure}

Seeding the pre-supernova profile of the star with realistic perturbations may be important to 
understanding how fluid instabilities later erupt and mix the star during the explosion.  Massive 
stars usually develop convective zones prior to exploding as SNe\cite{woosley2002}{}.  
Multidimensional stellar evolution models suggest that the fluid inside the convective regions can 
be highly turbulent\cite{porter2000,arnett2011}{}.  However, in lieu of the 3D stellar evolution 
calculations necessary to produce such perturbations from first principles, multidimensional 
simulations are usually just seeded with random perturbations.  In reality, if the star is convective 
and the fluid in those zones is turbulent\cite{davidson2004}{}, a better approach is to imprint the 
multidimensional profiles with velocity perturbations with a Kolmogorov energy spectrum \cite{frisch1995}{}. The scheme\cite{chen_phd} is the first attempt to model the initial 
perturbations based on a more realistic setup. Figure~\ref{perb2} shows the initial velocity 
perturbation by using the turbulent perturbation scheme. 
We seed initial perturbations to trigger the fluid 
instabilities on multidimensional simulations so we can study how they evolve with their surroundings.  
When the fluid instabilities start to evolve nonlinearly, the initial imprint of perturbation would be 
smeared out.  The random perturbations and turbulent perturbations then give consistent results.  Depending on the nature of the problems, the random perturbations might take a longer time to evolve 
the fluid instabilities into turbulence because more relaxation time is required.

\subsection{GR Correction}
\label{gr_sec}
In the cases of very massive stars ($\ge 1000\,\Msun$), the general relativity (GR) effect starts to play a role
in the stellar evolution. First, we consider the hydrostatic equilibrium 
due to the effects of GR, then derive GR-correction terms for Newtonian gravity. 
The correction term would be applied to the monopole-type of gravity calculation. 

The formulae of GR-correction here are based on [\refcite{kippen1990}]. For detailed physics, please refer 
to [\refcite{grbk2}]. In a strong gravitational field, Einstein field equations are required to 
describe the gravity:
\begin{equation}\label{field}
R_{ik}-\frac{1}{2}g_{ik}R=\frac{\kappa}{c^{2}}T_{ik}, \quad
\kappa=\frac{8\pi G}{c^{2}},
\end{equation} 
where $R_{ik}$ is the Ricci tensor, $g_{ik}$ is the metric tensor, $R$ is the Riemann curvature, $c$ is 
the speed of light, and $G$ is the gravitational constant. For ideal gas, the energy momentum tensor
$T_{ik}$ has the non-vanishing components $T_{00}$ = $\varrho c^2$ , $T_{11}$ = $T_{22}$ = $T_{33}$ = $P$ 
($\varrho$ contains rest mass and energy density; $P$ is pressure). We are interested in a spherically symmetric mass 
distribution. The metric in a spherical coordinate $(r, \vartheta, \varphi)$ has the 
general form
\begin{equation}\label{metric}
  ds^{2} = e^{\nu}c^{2}dt^{2}-e^{\lambda}dr^{2}-r^{2}(d\vartheta^{2}+\sin^{2}
  \vartheta d\varphi) ,
\end{equation} 
with $\nu = \nu(r)$, $\lambda = \lambda(r)$. Now insert $T_{ik}$ and $ds$ into Equation~(\ref{field}), 
then field equations can be reduced to three ordinary differential equations:
\begin{equation}\label{diff1}
   \frac{\kappa P}{c^{2}} =
   e^{-\lambda}(\frac{\nu^{\prime}}{r}+\frac{1}{r^{2}})-\frac{1}{r^{2}}
   ,
\end{equation}
\begin{equation}\label{diff2}
  \frac{\kappa P}{c^{2}} =
  \frac{1}{2}e^{-\lambda}(\nu^{\prime\prime}+\frac{1}{2}{\nu^{\prime}}^{2}+\frac{\nu^
    {\prime}-\lambda^{\prime}}{r}
   -\frac{\nu^{\prime}\lambda^{\prime}}{2}) ,
\end{equation}
\begin{equation}\label{diff3}
  \kappa \varrho =
  e^{-\lambda}(\frac{\lambda^{\prime}}{r}-\frac{1}{r^{2}})+\frac{1}{r^{2}} ,
\end{equation} 
where primes means 
the derivatives with respect to $r$. After multiplying with $4\pi r^2$, Equation~(\ref{diff3}) can 
be integrated and yields
\begin{equation}\label{gmass1}
  \kappa m = 4\pi r (1-e^{-\lambda}) ;
\end{equation}
$m$ is called  the ``gravitational mass'' inside r defined as
\begin{equation}\label{gmass2}
  m = \int_{0}^{r}4\pi r^{2}  \varrho dr .
\end{equation}      
For $r = R$, $m$ becomes the total mass $M$ of the star. $M$ here contains both the rest mass and 
energy divided by $c^2$. So the $\varrho = \varrho_0 +U/c^2$ contains the energy density $U$ and rest 
mass density $\varrho_0$. Differentiation of Equation~(\ref{diff1}) with respect to $r$ gives $P = P^{\prime}(\lambda,\lambda^{\prime},
\nu,\nu^{\prime},r)$, where $\lambda,\lambda^{\prime},\nu,\nu^{\prime}$  can be eliminated by Equations~(\ref{diff1}), 
(\ref{diff2}), (\ref{diff3}). Finally we obtain the Tolman--Oppenheinmer--Volkoff (TOV) equation for 
hydrostatic equilibrium in general relativity\cite{kippen1990}{}:
\begin{equation}\label{tov}
  \frac{dP}{dr} = -\frac{Gm}{r^{2}}\varrho (1+\frac{P}{\varrho
    c^{2}})(1+\frac{4\pi r^3 P}{m c^{2}}) (1-\frac{2Gm}{r c^{2}})^{-1} .
\end{equation}                  
For the Newtonian case $c^2 \rightarrow  \infty $, it reverts to the usual form,
\begin{equation}\label{newton}
  \frac{dP}{dr} = -\frac{Gm}{r^{2}}\varrho  .
\end{equation}
Now we take effective monopole gravity as
\begin{equation}\label{tov2}                                                      
\tilde{g} = -\frac{Gm}{r^{2}} (1+\frac{P}{\varrho
  c^{2}})(1+\frac{4\pi r^3 P}{m c^{2}}) (1-\frac{2Gm}{r c^{2}})^{-1}  .
\end{equation}                                 
For general situations, we neglect the $U/c^2$ and potential energy in $m$ because they are usually
much smaller than $\varrho_0$. Only when $T$ reaches $10^{13}\,\K$ (${\rm K}T \approx m_{p} c^2$,
 $m_{\rm p}$ is proton mass) does it start to make a difference. Equation~(\ref{tov2}) can be expressed as
\begin{equation}\label{tov3}                                                      
  \tilde{g} = -\frac{GM_{enc}}{r^{2}} (1+\frac{P}{\varrho
    c^{2}})(1+\frac{4\pi r^3 P}{M_{enc} c^{2}}) (1-\frac{2GM_{enc}}{r c^{2}})^{-1} ,
\end{equation}                                              
where $M_{enc}$ is the mass enclosure within $r$. Post-Newtonian correction of gravity is important 
for SNe from super massive stars, which will be discussed in \S~\ref{sn_gsn_section}.

\subsection{Resolving the Explosion}
 \label{resolve_sec}
In addition to implementing relevant physics for \CASTRO, care must be taken to determine the 
resolution of multidimensional simulations required to resolve the most important physical scales 
and yield consistent results, given the computational resources that are available.  We provide a 
systematic approach for finding this resolution for multidimensional stellar explosions.  

Simulations that include nuclear burning, which governs nucleosynthesis and the energetics of the 
explosion, are very different from purely hydrodynamical models because of the more stringent
resolution required to resolve the scales of nuclear burning and the onset of fluid instabilities in 
the simulations.  Because energy generation rates due to burning are very sensitive to temperature, 
errors in these rates as well as in nucleosynthesis can arise in zones that are not fully resolved.  
We determine the optimal resolution with a grid of 1D models in \CASTRO{}.  Beginning with a
crude resolution, we evolve the pre-supernova star and its explosion until all burning is complete 
and then calculate the total energy of the supernova, which is the sum of the gravitational energy, 
internal energy, and kinetic energy.  We then repeat the calculation with the same setup but with 
a finer resolution and again calculate the total energy of the explosion.  We repeat this process until the 
total energy is converged.  The time scales of burning (${\rm d}t_{\rm b}$) and hydrodynamics 
(${\rm d}t_{\rm h}$) can be very disparate, so we adopt time steps of 
$ min({\rm d}t_{\rm h},{\rm d}t_{\rm b})$ in our simulations, where 
${\rm d}t_{\rm h}=\frac{{\rm d}x}{c_{\rm s}+|v|}$; 
${\rm d}x$ is the grid resolution, $c_s$ is the local sound speed, $v$ is the fluid velocity, and 
the time scale for burning is ${\rm d}t_{\rm b}$, which is determined by both the energy generation 
rate and the rate of change of the abundances.

For simulating a thermonuclear SN, the spacial resolutions of $10^8\,\cm$ are usually
needed to fully resolve nuclear burning.  However, the star can have a radius of up to several $10^{14}\,\cm$.  This large dynamical range (10$^6$) 
makes it impractical to simulate the entire star at once
while fully resolving all relevant physical processes.  When the shock launches from the center of the star,
the shock's traveling time scale is about a few days, which is much shorter than the Kelvin--Helmholtz time
scale of the stars, about several million years.  We can assume that when the shock propagates inside the 
star, the stellar evolution of the outer envelope is frozen.  This allows us to trace the shock propagation without 
considering the overall stellar evolution.  Instead beginning simulations with a coordinate mesh that encloses just 
the core of the star with zones that are fine enough to resolve explosive burning.  We then halt the 
simulation as the SN shock approaches the grid boundaries, uniformly expand the simulation domain, and then restart the 
calculation.  In each expansion we retain the same number of grids.  Although 
the resolution decreases after each expansion, it does not affect the results at later times because burning 
is complete before the first expansion and emergent fluid instabilities are well resolved in later expansions.  
These uniform expansions are repeated until the fluid instabilities cease to evolve.  

Most stellar explosion problems need to deal with a large dynamic scale such as the case discussed here.  
It is computationally inefficient to simulate the entire star with a sufficient resolution.  Because the 
time scale of the explosion is much shorter than the dynamic time of stars, we can only follow the evolution 
of the shock by starting from the center of the star and tracing it until the shock breaks out of the stellar surface.

\subsection{Parallel Performance of \CASTRO}
\label{castro_mpi_sec}
\begin{figure}[ht]
\begin{center} 
\subfigure[{\bf Hopper} Supercomputer]{\label{scaling_a_fig}\includegraphics[width=0.45\textwidth]{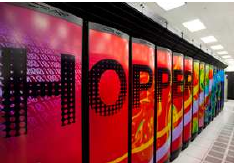}}                
\subfigure[\CASTRO{} scaling]{\label{scaling_b_fig} \includegraphics[width=0.45\textwidth]{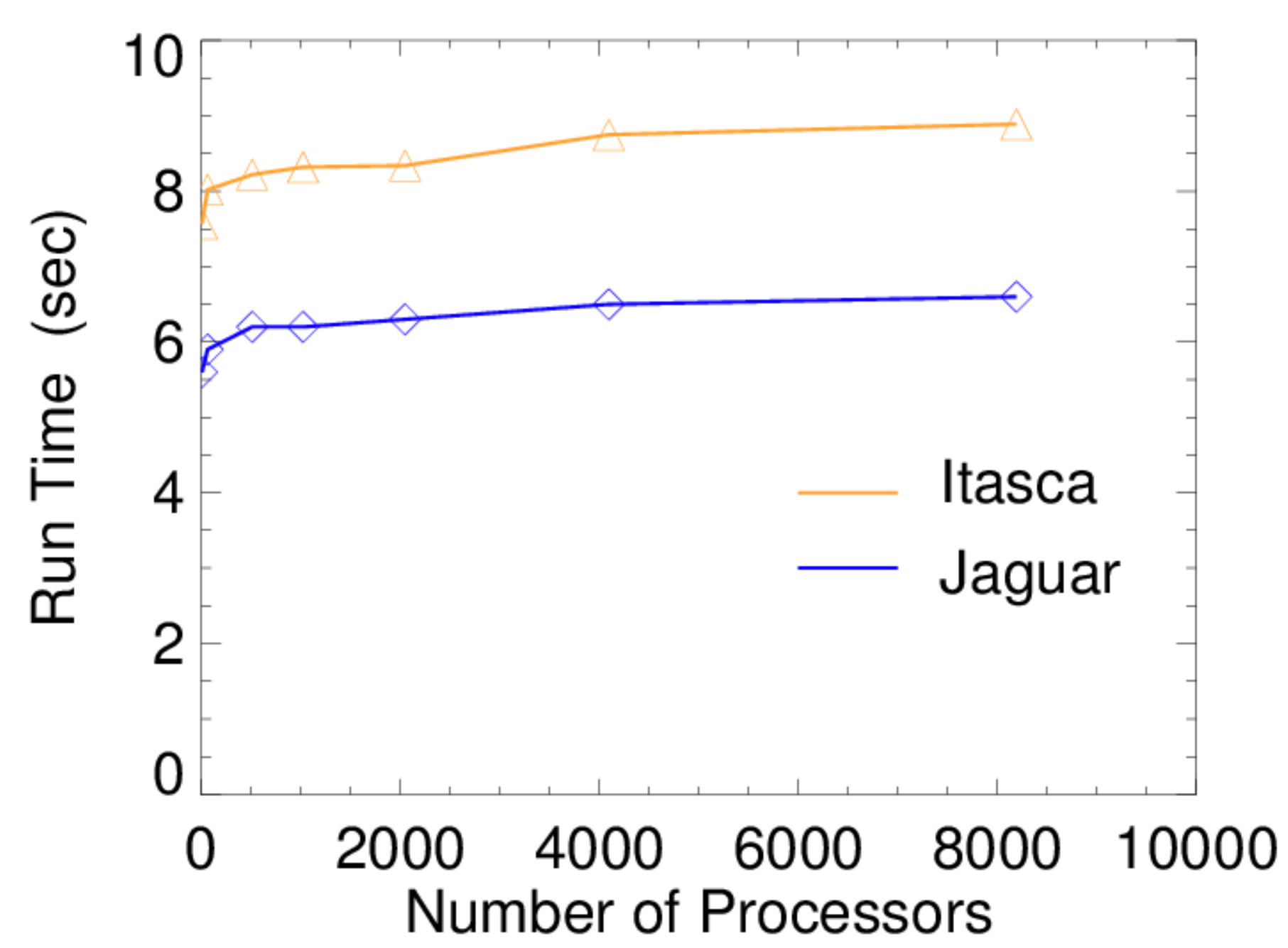}}   
\caption[\CASTRO{} scaling]{(a) Hopper: \CASTRO{} runs on some of the fastest supercomputers in the world, such as 
{\bf Hopper}  located at the Lawrence Berkeley Lab (Credit: NERSC website). 
(b) The weak scaling of \CASTRO{} on Itasca and Jaguar: The shock tube problem is used for 
the benchmark in the scaling of \CASTRO{}, and the number of processors is scaled to 
the load of the job. The symbols are the data from our results, for the case of perfect scaling, 
the curves should be flat. (Jaguar data provided by Ann Almgren and Andy Nonaka)}
\end{center}
\end{figure}

A multi-D SN simulation may need from hundreds of  thousands to millions of CPU hours to run. 
The parallel efficiency of the code becomes a very critical issue.It is usually good to find out 
how well the code parallels the jobs before we start burning tons of CPU hours.  
To understand the parallel efficiency of \CASTRO, a weak scaling study is performed, 
so that for each run there is exactly one 64$^3$ grid per processor. We run the Sod problem on 
$32 (1024\times256\times256)$, $256 (2048\times512\times512)$, $512 (2048\times1024\times512)$,
$1024 (2048\times1024\times1024)$, and $8192 (4096\times2048\times2048)$ CPU on Itasca at 
the Minnesota Supercomputing Institute (MSI); the grid information is 
inside the parentheses. Our collaborators also perform weak scaling tests on the Jaguar at the Oak Ridge Leadership 
Computing Facility, which runs white dwarf 3D problems on 8, 64, 512, 1024,  2048, 4096 and 8192 processors. 
Figure~\ref{scaling_b_fig} shows the weak scaling of \CASTRO{} on Itasca and Jaguar. 
For these scaling tests, we use only MPI-based parallelism with non-AMR grids.  The results 
suggest \CASTRO{}  demonstrates a satisfying scaling performance within the number of CPU 
between $32-8192$ on both supercomputers. The scaling behavior of \CASTRO{} may depend on the 
calculations, especially while using AMR.

\subsection{Visualization with \VISIT}
\label{visit_sec}
\CASTRO{} uses Boxlib as its output format. Depending on the dimensionality and resolution 
of simulations,  the \CASTRO{} outputs  can be as massive as hundreds of Gigabytes. Analyzing
and visualizing such data sets becomes technically challenging.  We visualize and analyze the data 
generated from \CASTRO{} by using  custom software, \VISIT{} \cite{visit}{}, an interactive parallel 
visualization and graphical analysis tool. \VISIT{} is developed by the DOE, Advanced Simulation and 
Computing Initiative (ASCI), and it is designed to visualize and analyze the results from large-scale 
simulations. \VISIT{} contains a rich set of visualization features, and users can implement their tailored 
functions on \VISIT. Users can also animate visualizations through time, manipulate them, and save 
the images in several different formats. For our simulations, we usually use a pseudocolor 
plot for 2D visualization and a contour plot or volume plot for 3D visualization.  The pseudocolor 
plot maps the physical quantities to colors on the same planar and generates 2D images. The contour 
maps 3D structures onto 2D iso-surfaces, and the volume plots fill 3D 
volume with colors based on their magnitude. Visualizing data also requires the supercomputing 
resources, especially for storage and memory. Most of SN images in this review were generated
using \VISIT.

\section{Pop~III Supernovae - Explosions}
\label{sn_result}
 In this section, we present the recent results of the Pop~III supernovae based on [\refcite{chen_phd}].
 These SNe came from the thermonuclear SNe of very massive Pop~III stars above $80\,\Msun$.  
 We will discuss the physics of the formation of these fluid instabilities during the SN explosions. 
  
\subsection{Fate of Very Massive Stars I $\mathbf{(80\,M_{\odot}\,\leq\,M_*\,<\,150\,M_{\odot})}$} 
 After the central carbon burning, the massive stars over $80\,\Msun$ become unstable because part of 
energetic photons start to convert into $e^-/e^+$ inside their core.  The removal of radiative pressure 
softens the adiabatic index $\gamma_{\rm a}$ below $4/3$. Central temperatures 
start to oscillate with a period about the dynamic time scale of $500\,\sec$. However these oscillations in 
temperatures do not send shock into the envelope to produce any visible outburst. The star still goes 
through all the advanced burnings before it dies as a CCSN.  If the mass of the star is close to $100\,\Msun$,  central 
temperatures again fluctuate due to pair-instabilities right after carbon burning. The 
amplitude of oscillation becomes larger.  Several shocks incidentally are sent out from the core 
before the stars die as CCSNe. The energy of a pulse is about $\Ep{50}\,\erg$ 
([\refcite{woosley2007}]; Woosley, priv. comm.), while the typical binding energy for the hydrogen 
envelope of such massive stars is less than $\Ep{49}\,\erg$. These shocks are inadequate to blow up the 
entire star, but they are strong enough to eject several solar masses from the stellar envelope. 

\begin{figure}[ht]
\begin{center}
\includegraphics[width = \columnwidth]{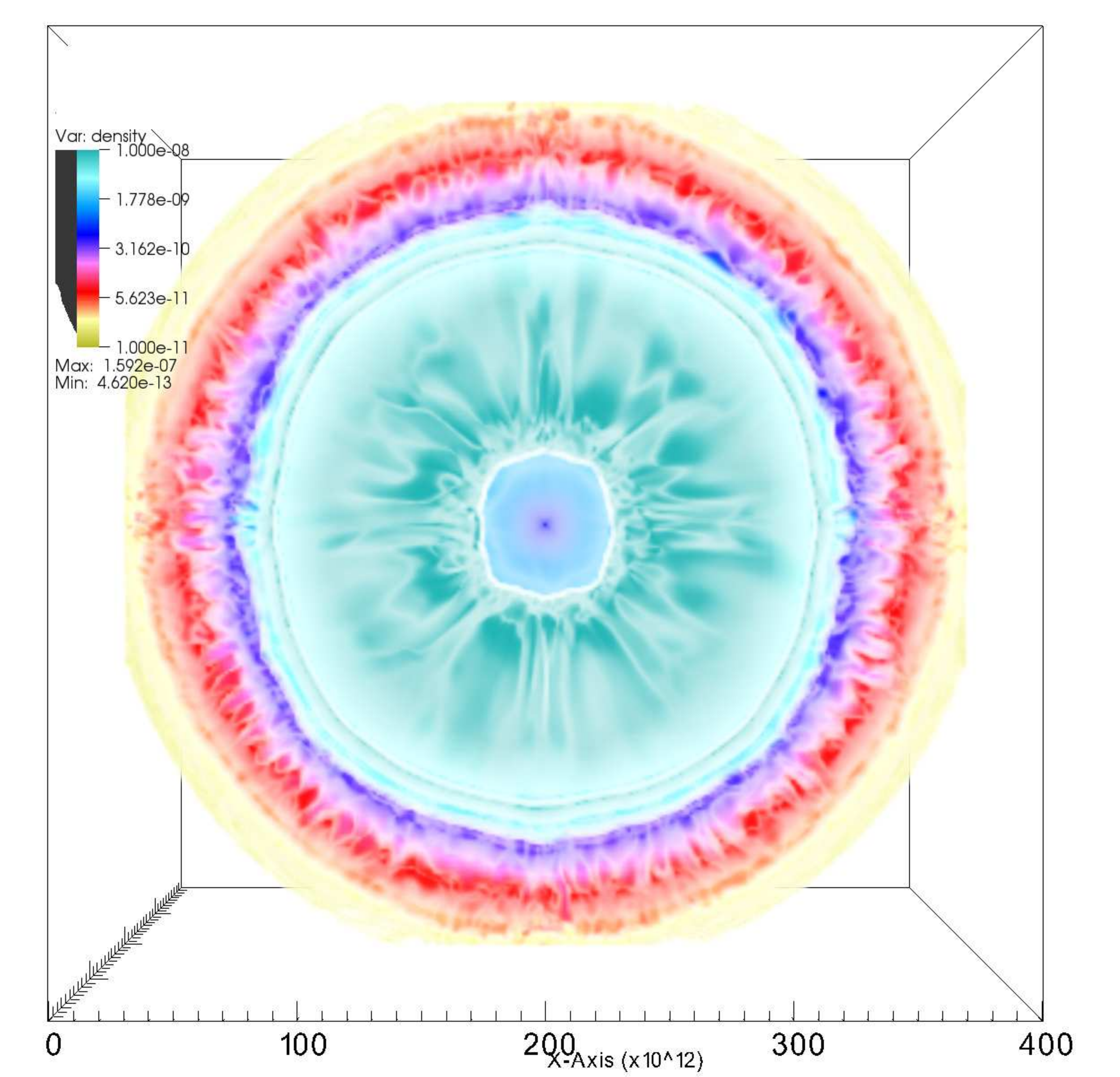}
\caption[]{Collision of PPSN shells:  Colors show the densities of SN ejecta. Many fluid instabilities 
occur in the purple-red regions where the shells collided. Because part of the kinetic ejecta is converted 
into thermal energy, this kind of collision can result in a strong  emission of thermal radiation.     \label{ppsn3d}}
\end{center}
\end{figure}

\begin{figure}[ht]
\begin{center}
\includegraphics[width=\columnwidth]{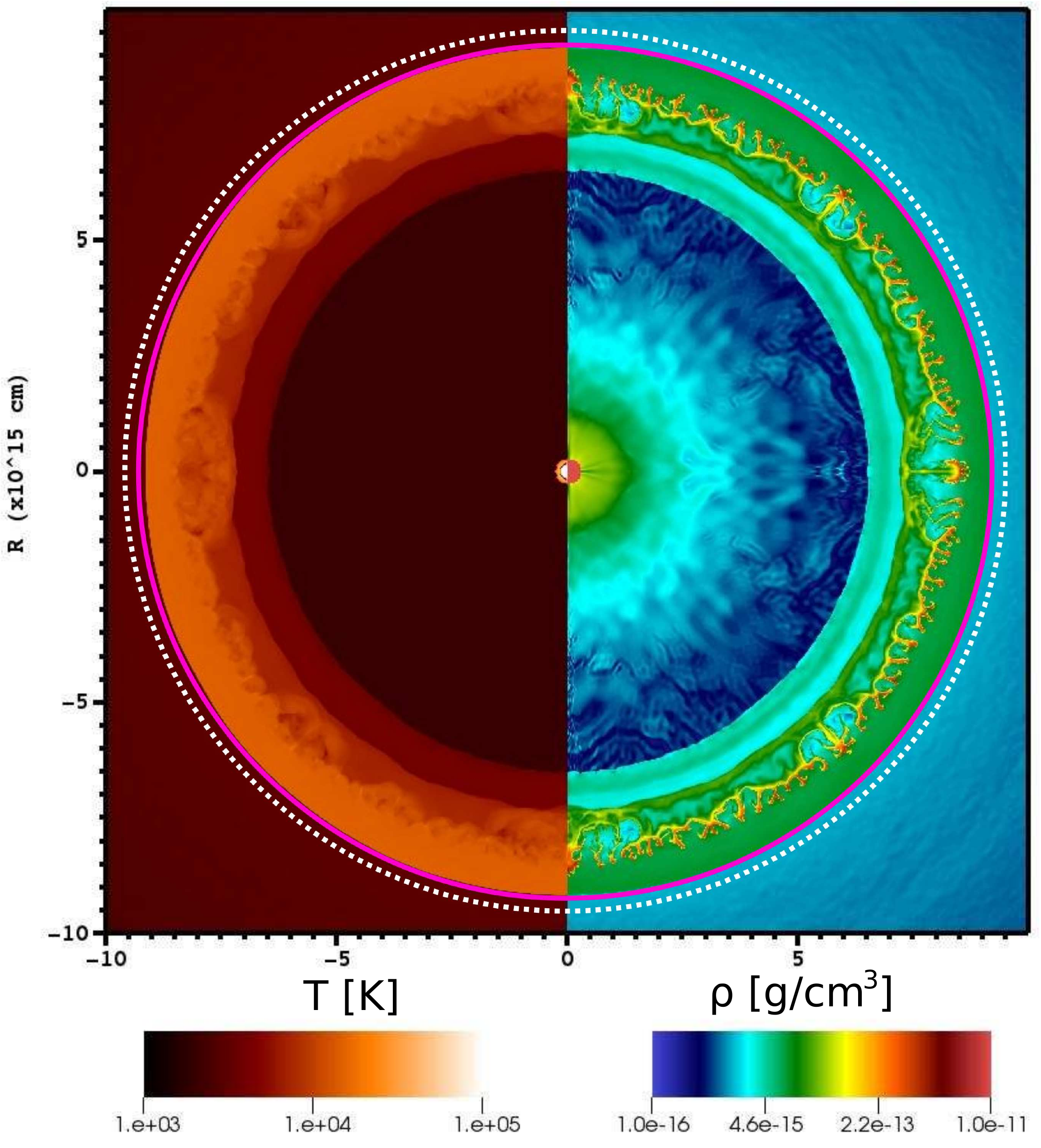} 
\end{center}
\caption[]{Density and temperature of PPSN: The pink dashed line shows the location of the 
shock front, and the white dashed line shows the photon sphere of star $\tau \sim 2/3$. 
At this time, the shock is about to break the photosphere of the star; the thermal emission of 
ejecta results in a very luminous optical transit.
\label{ppsn_collision}}
\end{figure}

We have performed the first 2D/3D simulations of the pulsational pair-instability supernovae 
with \CASTRO{}. In our 2D simulation of a  PPSNe of a $110\,\Msun$ star, we  found fluid instabilities occurred during 
the fallback of ejecta and the collisions of ejected shells. Fallback of unsuccessful ejected shells caused 
minor fluid instabilities that did not result in much mixing. However, the catastrophic collisions of pulses 
produce many fluid instabilities.  The heavy elements ejected from the star are mainly \Ox{} and \Cx. 
The latter outbursts are more energetic than the earlier ones, that leads to the collision of ejecta.
When the ejecta from different eruptions collide, significant mixing 
is caused by the fluid instabilities as shown in Figure~\ref{ppsn3d}. Collision of ejecta efficiently converts their kinetic energy into thermal 
energy that releases in the form of photons.  The clumped structure caused by fluid instabilities may trap the 
thermal photons during the collision and affect the observational luminosity. The mixed region is very close to the photo-sphere of PPSNe, as shown in Figure~\ref{ppsn_collision} and 
potentially alters their observational signatures. The mixture of the ejecta can also affect 
the spectra by altering the order in which emission and absorption lines of particular 
elements appear in the spectra over time. The radiation transport is required for modeling 
such a complex process of radiation coupled with flow of gas before 
obtaining the light curves and spectra for these transients.
We expect the mixing can intensify because the radiation cooling of
clumps is amplified by the growth of fluid instabilities.

\subsection{Fate of Very Massive Stars II ({ $\mathbf{{150}\,M_{\odot}\,\leq\,M_*\,\leq\,260\,M_{\odot}}$}) }
Pop~III stars with initial masses of $150-260\,\Msun$ develop oxygen cores of 
$\gtrsim$ $50\,\Msun{}$ after central carbon burning \cite{barkat1967,glatzel1985,heger2002,heger2010}{}.  
At this point, the core reaches sufficiently high temperatures ($\sim \Ep9\,\K$) and at relatively low 
densities ($\sim\Ep6\,\gcc$) to favor the creation of $e^-/e^+$ (high-entropy hot
plasma).  The pressure-supporting photons turn into the rest masses
for pairs and soften the adiabatic index $\gamad$ of the gas
below a critical value of $\nicefrac43$, which  causes a dynamical
instability and triggers rapid contraction of the core.  During
contraction, core temperatures and densities swiftly rise, and oxygen
and silicon ignite, burning rapidly.  This reverses the preceding
contraction (enough entropy is generated so the equation of state
leaves the regime of pair instability), and a shock forms at the outer
edge of the core.  This thermonuclear explosion, known as a
pair-instability supernova (PSN), completely disrupts the star with
explosion energies of up to $\Ep{53}\,\erg$, leaving no compact
remnant and producing up to $50\,\Msun$ of \Ni{}\cite{heger2002,kasen2011}{}.

\begin{figure}[ht]
\begin{center}
\includegraphics[width=\textwidth]{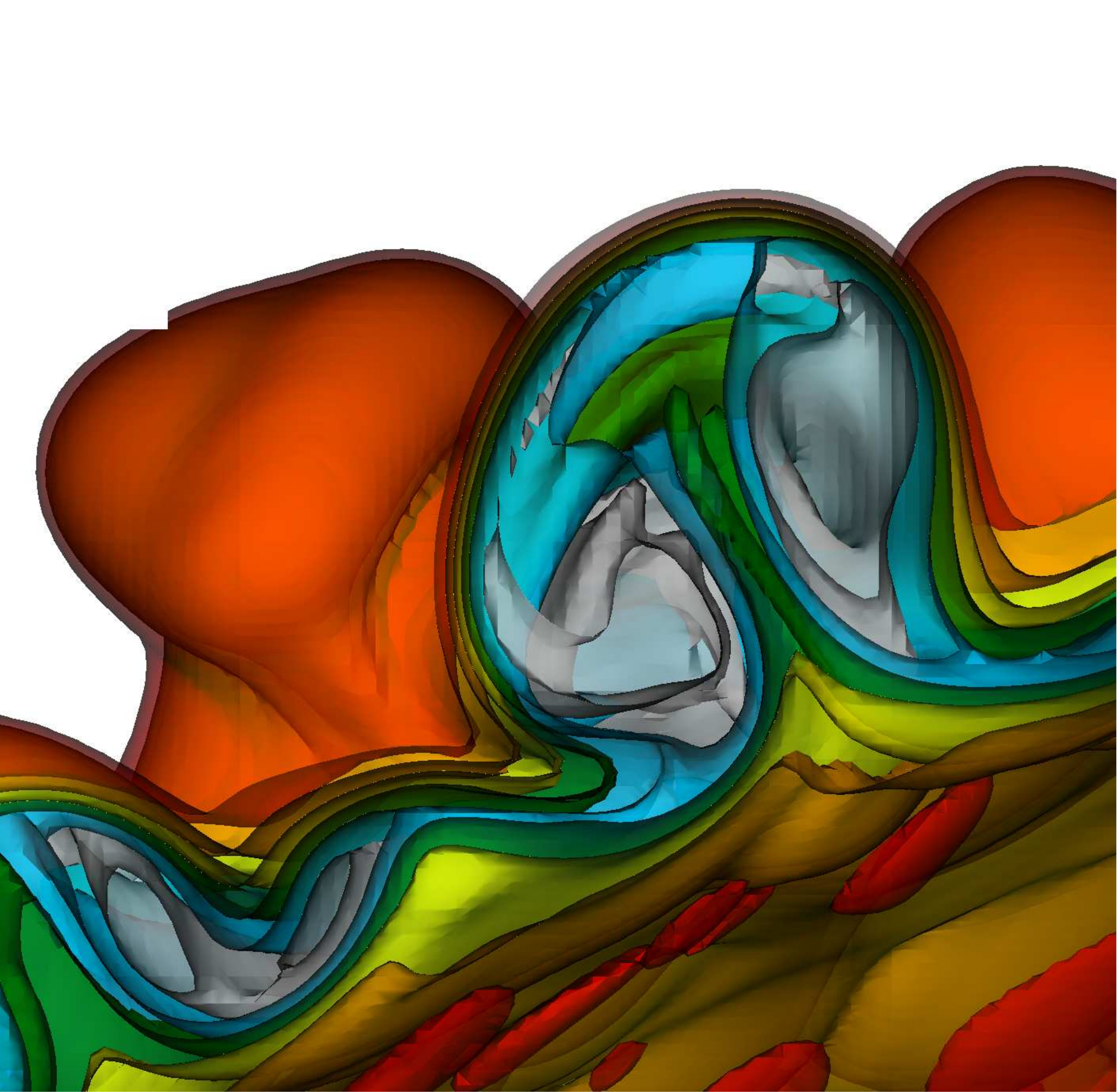}
\caption[]{3D fluid instabilities inside a PSN: Colors show the carbon abundance.  Fluid instabilities driven by 
nuclear burning occurred at the very early stage of explosion. Such fluid instabilities only cause a minor 
mixing by dredging  up a little of material.
    \label{3d_blue}}
\end{center}
 \end{figure}

\begin{figure}[ht]
\begin{center}
\includegraphics[width = \columnwidth]{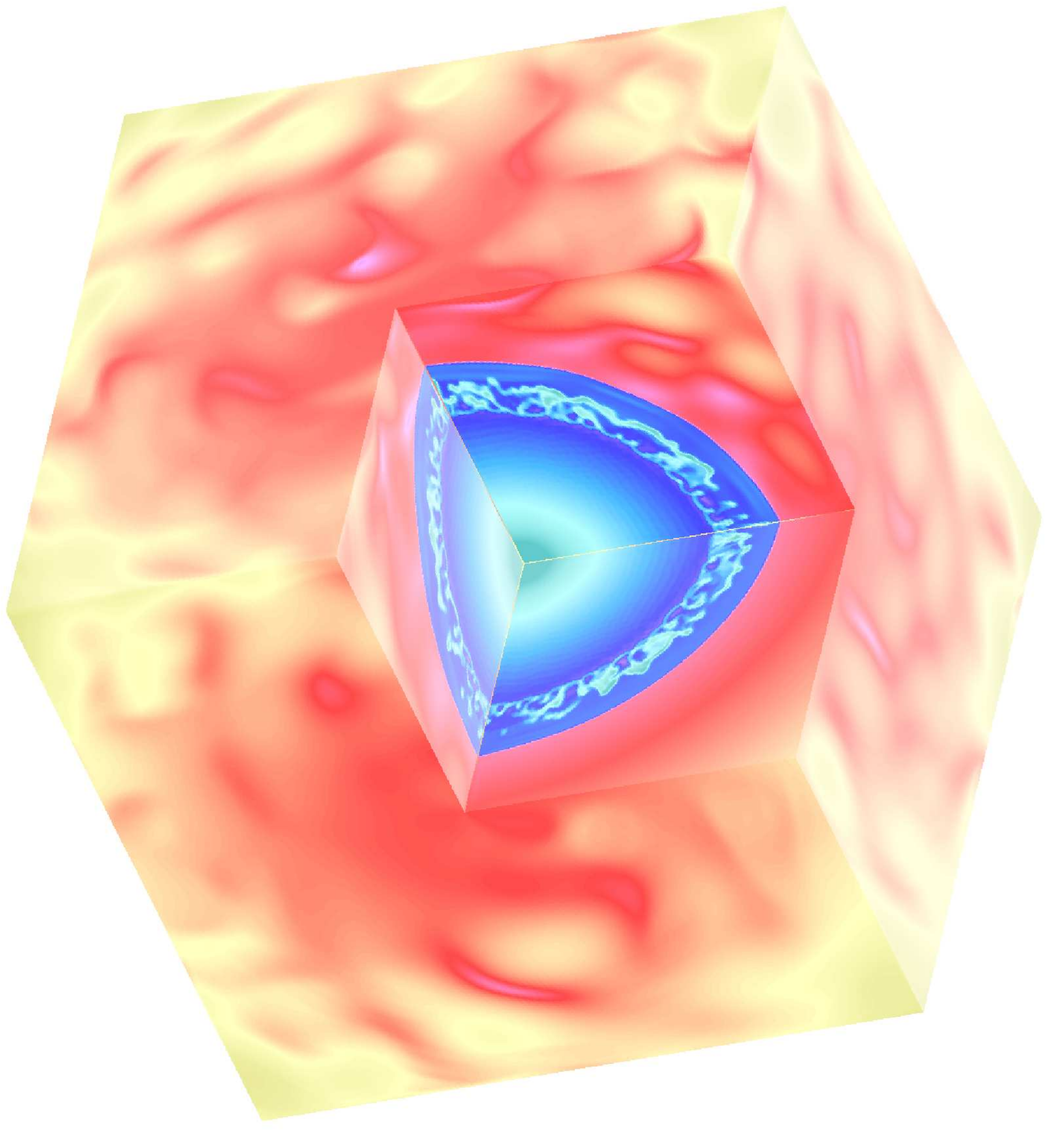}
\caption[]{ 3D PSN from a red supergiant: Fluid instabilities driven by the reverse-shock are sufficient to  
mix up the SN ejecta in a large scale.   \label{3d_red}}
\end{center}
\end{figure}

Multidimensional simulations suggest that fluid instabilities occur  at the different phases of explosion: 
collapse, explosive burning, and shock propagation. The particular phase depends on the pre-SNe progenitors. 
For blue supergiants, the fluid instabilities driven by nuclear burning occur at the very beginning of explosion. 
Such instabilities only lead a minor mixing at the edges of the oxygen-burning shells 
due to a short growth time, $\leq 100\,\sec$, as shown in Figure~\ref{3d_blue}. 
The red supergiants show a strong mixing which breaks 
the density shells of SN ejecta.  Because when the shock enters into the hydrogen 
envelope of red supergiants, it is decelerated by snowplowing mass that grows the Rayleigh-Taylor instabilities 
\cite{chan1961}{}. Figure~\ref{3d_red} shows a visible mixing due to Rayleigh-Taylor instabilities inside 
a red supergiant. However, mixing inside PSN is unable to dredge up \Ni{} before the shock breakout.

\subsection{Fate of Extremely Massive Stars III ($\mathbf{M_* \gg 100\,M_{\odot}}$)}
\label{sn_gsn_section}
Results from observational and theoretical studies [\refcite{kormendy1995,ferra2000,ferra2005,gebh2000,beif2012,mcco2013}] 
suggest that a supermassive black hole (SMBH) resides in each galaxy. These SMBHs play an important role in the 
evolution of the Universe through their feedback. Like giant monsters, they swallow nearby stars and gas, and spit out strong x-rays  
and powerful jets \cite{rees1984,matte2005} that impact scales from galactic star formation to host galaxy clusters. 
Quasars \cite{fan2002,fan2006} detected at the redshift of $z \ge 6$  suggest that SMBHs had already formed when the Universe was 
only several hundred million years old. But how did SMBHs form in such a short time?  

\begin{figure}[ht]
\begin{center}
\includegraphics[width = \columnwidth]{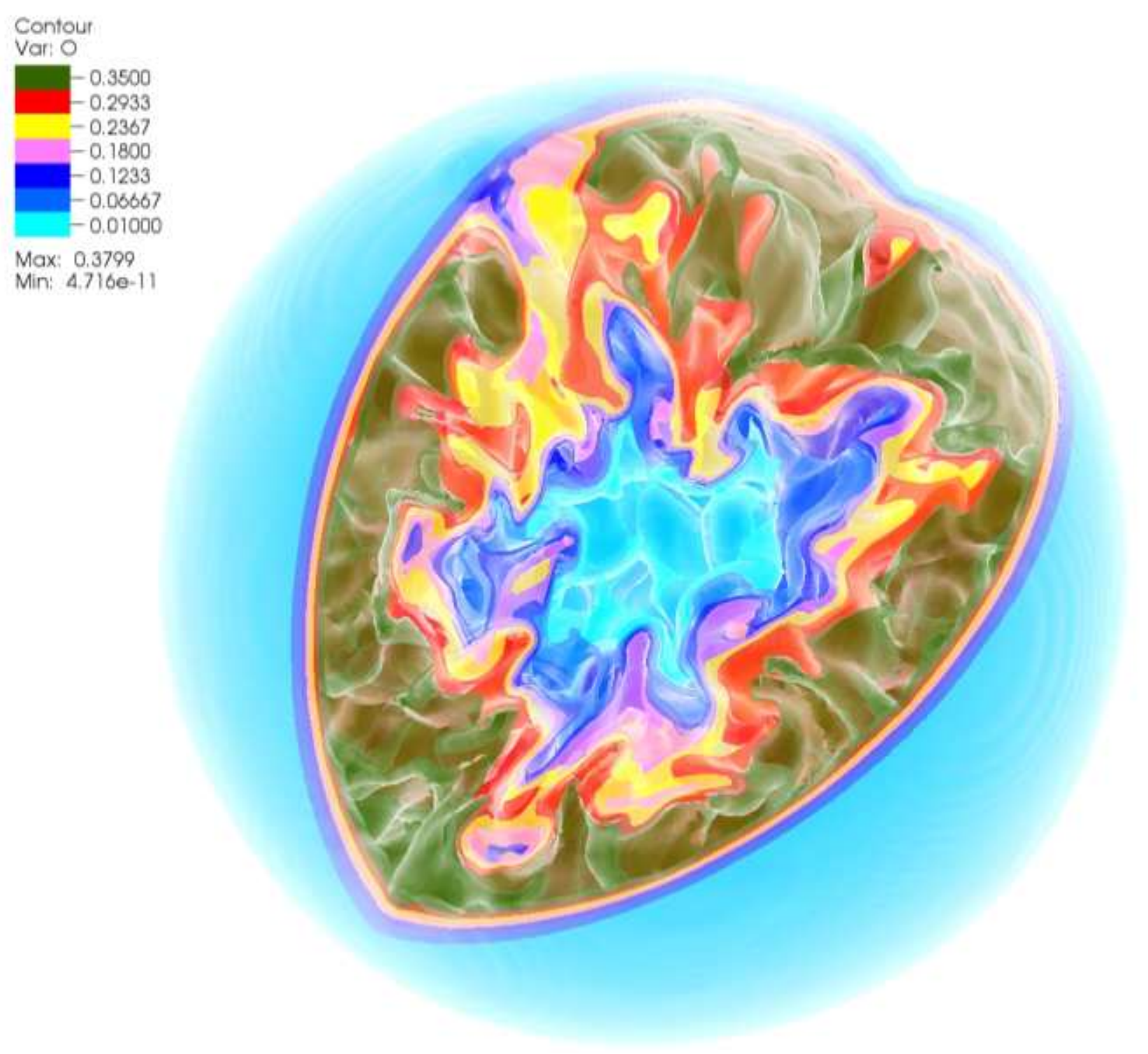}
\caption[]{ An exploding supermassive star of $55,500\,\Msun$: Colors show the oxygen mass fraction of the inner core of the star. Many fluid instabilities have occurred right after the bounce of the core. \label{gsn}}
\end{center}
\end{figure}

Models for the formation of SMBHs in the early Universe have been extensively 
discussed by many authors: [\refcite{loeb1994,madau2001,bromm2003,bege2006,johnson2007,bromm2011}].
[\refcite{rees1984}] first pointed out the pathways of forming SMBHs. One of the possibilities is through 
the channel of super massive stars (SMS) with masses $\ge 10,000$ \Msun. They might 
form in the center of the first galaxies through atomic hydrogen cooling [\refcite{johnson2012}]. If SMS
could form in the early Universe, they could facilitate SMBH formation by providing 
promising seeds. Although the mechanism of SMS formation is not clear, the evolution of SMS 
has been studied by theorists [\refcite{fowler1966,wheeler1977,bond1984,carr1984,fuller1986,fryer2001,ohkubo2006}] for three decades.  Previous results of [\refcite{fryer2001,ohkubo2006}] 
suggest that non-rotating stars with initial masses over $300\,\Msun$ eventually die as black holes without supernova explosions.  
It is generally believed that the explosive burning is insufficient to revert the implosion because the SN shock is dissipated by the photo-disintegration 
of the heavy nuclei; thus, these stars eventually die as BHs without SN explosions. 
[\refcite{chen_phd}] found an unusual explosion of a SMS of $55,500\,\Msun$ that implies a narrow 
mass window for exploding SMS, called 
General-Relativity instability supernovae (GSNe). GSNe may be triggered by the general relativity instability that happens after 
central helium burning and leads to a runaway collapse of the core, eventually igniting the 
explosive helium burning and unbinding the star. The energy released from the burning is large enough to 
reverse the implosion into an explosion and unbind the SMS without leaving a compact remnant as shown 
in Figure~\ref{gsn}. 
Energy released from the GSN explosion is about $10^{55}\,\erg$, which is about $10,000$ times 
more energetic than is typical of supernovae. The main yields of SMS explosions are silicon 
and oxygen; only less than $1\,\Msun$ \Ni{} is made. The ejecta mixes due to the fluid instabilities driven 
by burning during the very early phase of the explosion. We list the characteristics of PPSNe, PSNe, and 
GSNe in Table~\ref{sn_comp}.

\begin{table}[ht]
   \tbl{Charactersitics of PPSNe, PSNe and GSNe}
   {\begin{tabular}{@{}lccc@{}} \toprule
Characteristic Property  &   PPSNe & PSNe &  GSNe   \\ \colrule
 Mass of Progenitor [$\Msun$] & 80 -150 & $150 - 260$     &  $55500\pm ??$                  \\
 Collapse Trigger & Pair Instabilities     & Pair Instabilities     &     GR Instabilities     \\
 Burning Driver &   \Ox{} & \Ox, \Si {}  &   \He{}               \\
 \Ni{} Production  [$\Msun$] & $\ll 1$  &   $0.1 - 50$ & $\ll 1$  \\
 Explosion Energy [$\erg$] & $1-100 \times 10^{49}$    & $1-100 \times 10^{51}$     &   $6-10\times10^{54}$ \\
 Fluid Instabilities  & Colliding Shells   &    Reverse Shock &  Burning   \\ 
   \botrule
\end{tabular} \label{sn_comp}}
   \end{table}

\subsection{Candidates for Superluminous Supernovae in the Early and Local Universe}
\begin{figure}[ht]
 \begin{center}
 \includegraphics[width=\columnwidth]{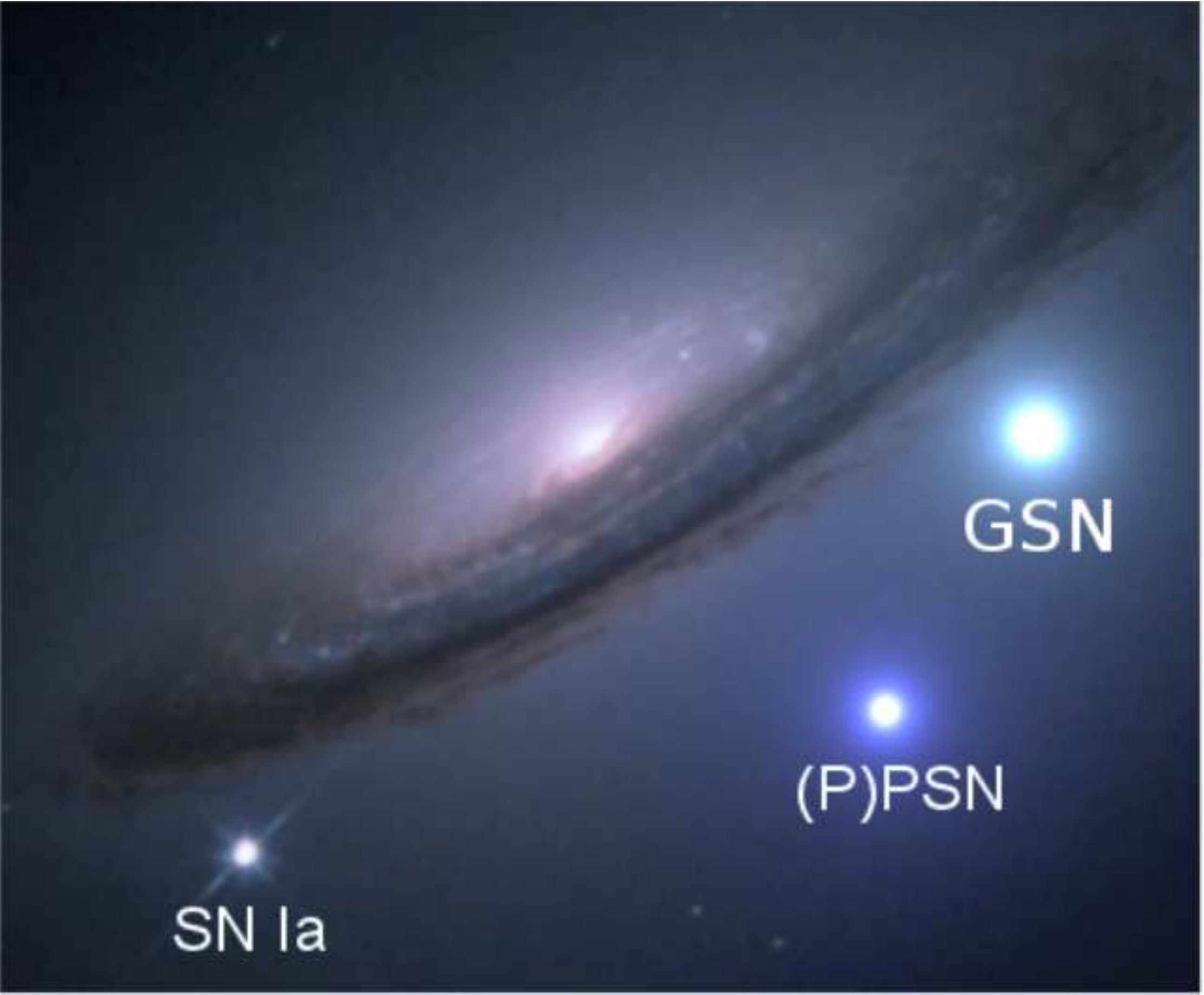}
 \caption[]{Pseudo observational 
 signatures of supernovae: Ia, PPSN, PSN, and GSN. Due to the enormous explosion energy or 
 large out-shining radius, (P)PSN or GSN can be $10-100$ times brighter than a type Ia SN.  (Original image credit: NASA/HST/High-z SN Search Team)\label{all_sn}}
 \end{center}
 \end{figure}

Because of the advancement of modern CCDs, the detection rates of SNe have rapidly increased. Large SN surveys, 
such as the Nearby Supernova Factory \cite{snf1, snf2} and the Palomar Transient Factory \cite{ptf1, ptf2}{}, have rapidly 
increased the volume of SN data and sharpened our understanding of SNe and their host environments.  
More and more supernovae defying our previous classifications have been found in the last decade; they have challenged our 
understanding of the SN progenitors, their explosion mechanisms, and their 
surrounding environments. One new type of SNe  found in recent observations is the superluminous SNe (SLSNe)\cite{galyam2012}{}, 
such as SNe 2006gy and 2007bi [\refcite{smith2007,galyam2009,past2010,quimby2007,quimby2011}]{}, which  shine an order of magnitude brighter 
than general SNe that have been well studied in the literature\cite{filip1997,smartt2009}{}. These SLSNe are relatively scarce, 
comprising less than $5\%$ of the total number of SNe that have been detected. They are usually found in galaxies with a lower
brightness, e.g., dwarf galaxies. The engines of SLSNe challenge our understanding of CCSNe. First, the luminosity of SNe can be
simply approximated in the form: $\propto 4\pi r^2 T^4$, where is $r$ is the radius of the photo-sphere, and $T$ is its effective 
surface temperature. If we assume the overall luminosity from the black body emission of hot ejecta, it requires either larger $r$ or $T$ to produce a 
more luminous SN. $r$ is determined when the hot ejecta becomes optically thin; then the photons start to stream freely. $T$
depends on the thermal energy of ejecta, which is directly related to the explosion energy. The duration of light curves is associated with 
the mass of ejecta determining the diffusion time scale and the size of the hot reservoir. PPSNe and PSNe are ideal candidates 
of SLSNe.  The collision shells of PPSNe  can generate very luminous transits \cite{woosley2007}{}. PSNe are also ideal candidates 
for SLSNe because of their huge explosion energy and massive \Ni{} production\cite{kasen2011}{}. Radioactive isotopes \Ni{}  can 
decay into \Co{} then \Fe{}, which releases much energy to lift up the light curve of SNe. 
It is promising that future large space and ground observatories, such as the James Webb 
Space Telescope (JWST), may be able to directly detect these SNe from Pop~III stars.
Although the 
gigantic explosions make the GSNe also a viable candidate, its huge mas makes the transit 
time of SNe last for several decades. Observation of GSNe become very difficult.  Figure~\ref{all_sn} shows an artificial  image of the observational signatures
of PPSNe, PSNe, and GSNe, which can be of one or two orders of magnitude brighter than normal SNe.

\section{Supernova Feedback with {\GADGET}}
\label{feed_gadget}
If the first stars were massive and died as SNe, their energetics and synthesized 
metals must have returned to the early Universe. An important question arises: How does the stellar 
feedback of the first stars impact the early Universe and how do we model such feedback?
In this section, we describe our computational approaches of feedback simulations by introducing the 
features of \GADGET{} and additional physics modules that we use for feedback simulations. 
We first introduce the hydrodynamics and gravity of SPH of \GADGET{} in \S~\ref{sph_sec}.
The cooling and chemical network of the primordial gas is discussed in \S~\ref{cooling_sec}. 
Since the star formation in the context of cosmological simulations cannot be modeled from 
first principles, we explain the sink particle approach for star formation in \S~\ref{sink_sec}. 
Once the first stars form in the simulation, they start to emit UV photons. Most of these stars 
would die as SNe. Under the context of cosmological simulations, we discuss the radiation transfer of UV 
photons in \S~\ref{ray_sec} and the supernovae feedback in \S~\ref{sn_sec}. Finally, we present 
a scaling performance of \GADGET{} in \S~\ref{gadget_sec}.

\subsection{Smoothed Particle Hydrodynamics}
\label{sph_sec}
$\GADGET{}$\cite{springel2005} ({\bf GA}laxies with {\bf D}ark matter and {\bf G}as int{\bf E}rac{\bf T})
is the main tool for our cosmological simulations. It is a well-tested, 
massively-parallel cosmological code that computes gravitational forces by using a 
tree algorithm and models gas dynamics by using smoothed particle 
hydrodynamics (SPH). We discuss the modified version of \GADGET{} including 
the relevant physics of the early Universe, such as star formation, radiative 
transfer, cooling, and chemistry.  Cosmological simulations
need to resolve the small-scale resolution under a huge domain. The SPH approach uses the 
Lagrangian coordinate instead of a spatial coordinate and is suitable for cosmological simulations. 
In addition to hydro and gravity, our simulations consider several feedback elements
from the first stars, e.g., radiation, supernova explosion, metal diffusion, et al. 
Major code development was done by Prof.~Volker Bromm and his group 
at the University of Texas.

Smoothed particle hydrodynamics\cite{monaghan1992} uses a mesh-free Lagrangian method 
by dividing the fluid into discrete elements called particles. Each particle has its own position ($r_i$), 
velocity ($v_i$), mass ($m_i$), and thermal dynamical properties, such as internal 
energy per unit mass ($u_i$). Additionally, each particle is given a physical size 
called smoothing length ($h$). The distribution of physical quantities 
inside a particle is determined by a kernel function ($W$). The most 
popular choices of kernel functions are Gaussian and cubic spline functions.
When each particle evolves with the local conditions, the 
smoothing length changes, so the spatial resolution of the fluid element  
becomes adaptive, which allows SPH to handle a large dynamic
scale and be suitable for cosmological simulations.
 $h$ of particles in higher-density regions becomes smaller because more particles 
 accumulate. SPH automatically increases the spatial resolution of simulations. The 
major disadvantages of SPH are in catching 
shock fronts and resolving the fluid instabilities because of
its artificial viscosity formulation, which injects the necessary entropy in shocks. 
The shock front becomes broadened over the smoothing scale, and true contact 
discontinuities cannot be resolved. However, SPH  are very suitable for simulating 
the growing structures due to gravity, and SPH adaptively resolves higher-density 
regions of halos, which are usually the domain of interest.

The cold dark matter is collisionless particles, and they interact with each 
other only through gravity. Hence gravity is the dominating force that drives the 
large-scale structure formation in the Universe, and its computation is the workhorse 
of any cosmological simulation. The long-range nature of gravity 
within a high dynamic range of structure formation problems makes the 
computation of gravitational forces very challenging. In \GADGET{}, the algorithm 
of computing gravitational forces employs the hierarchical multipole expansion called 
a tree algorithm. The method groups distant particles into larger cells, allowing their 
gravity to be accounted for by means of a single multipole force. 
For a group of N particles, the direct-summation approach needs N -1 partial 
forces per particle, but the gravitational force  using the tree method only requires 
about log N particle forces per particle. This greatly saves the computation cost. 
The most important characteristic of a 
gravitational tree code is the type of grouping employed. As a grouping algorithm, \GADGET{} 
uses the geometrical oct-tree \cite{bh1986} because of advantages in terms of memory 
consumption. The volume of the simulation is divided up into cubic cells in an oct-tree. 
Only neighboring particles are treated individually, but distant particles are grouped into a 
single cell. The oct-tree method significantly reduces the computation of pair 
interactions more than the method of direct N-body.

\subsection{Cooling and Chemistry Networks of Primordial Gas}
\label{cooling_sec}
Cooling of the gas plays an important role in the star formation. 
The dark matter collapses into halos and provides gravitational 
wells for the primordial star formation. The mass of the gas cloud must be
larger than its Jeans mass so the star formation can proceed. Cooling is an 
effective way to decrease the Jeans mass and trigger the star formation. 
The chemical cooling of the first star formation is relatively simple because 
no metals are available coolants at the time. 
 
According to [\refcite{bromm2004a}], the dominant coolant in the first star formation is molecular hydrogen. For the local Universe, the formation of {\Hm} occurs mainly at the surface of 
dust grains, where one hydrogen atom can be attached to the dust surface and combine 
with another hydrogen atom to form {\Hm}. There is no dust when Pop~III
stars form; the channel of {\Hm} through dust grain is unavailable. {\Hm} formation 
of primordial gas can only go through gas phase reactions. The simplest 
reaction is 
\be
\Ha + \Ha \longrightarrow \Hm + \gamma,
\ee
which occurs when one of the hydrogen atoms is in an electronic state. 
When the densities of hydrogen become high
enough, $n_\Ha\,\ge\,10^8\,\cc$, three-body formation of $\Hm$ becomes possible:
\be
\Ha + \Ha + \Ha \longrightarrow \Hm + \Ha,
\ee
\be
\Hm + \Ha + \Ha \longrightarrow \Hm + \Hm.
\ee
For the first star formation, the cloud collapses at the densities $n_\Ha\,\sim\,10^4\,\cc$. 
$\Hm$  is dominated by two sets of reactions: 
\be
\Ha + \ex \longrightarrow \Ha^- + \gamma,
\ee
\be
\Ha^- + \Ha  \longrightarrow \Hm + \ex.
\ee
This reaction involves the $\Ha^-$ ion as an intermediate state,
\be
\Ha + \Ha^+ \longrightarrow \Hm^+ + \gamma,
\ee
\be
\Hm^+ + \Ha  \longrightarrow \Hm + \Ha^+.
\ee
The second one involves the $\Ha^+$ ion as an intermediate state.
These two processes are denoted as the $\Ha^{-}$ pathway and the $\Hm^+$ pathway, respectively. 
The difference between the two pathways is that the $\Ha^-$ path forms $\Hm$ much faster than the 
$\Hm^+$ does, so the $\Ha^-$ pathway dominates the production of $\Hm$
in the gas phase. During the epoch of the first star formation, [\refcite{bromm2004a}] pointed out that 
molecular hydrogen fraction is $f_{\Hm}=10^{-3}\sim10^{-4}$ at minihalos and $f_{\Hm}\approx10^{-6}$
at the IGM. For given $\Hm$ abundances, density, and temperature, we are able to 
calculate the $\Hm$ cooling. The values of $\Hm$ cooling rates are not well-defined 
because of the uncertainties in the calculation of collisional de-excitation rates.

The cooling and chemistry network in our modified \GADGET{} is based on [\refcite{greif2010}] 
and include all relevant cooling mechanisms of primordial gas, such as H and He collisional 
ionization, excitation and recombination cooling, {\it bremsstrahlung}, 
and inverse Compton cooling; in addition, the collisional excitation cooling via $\Hm$ and HD is also taken into account. For $\Hm$ cooling, 
collisions with protons and electrons are explicitly included. The chemical network includes $\rm H, H^+, H^-, H_2, H_2^+, He, He^+, He^{++}$, and ${e^-}$, D, $\rm D^+$, and HD.

\subsection{Sink Particles}
\label{sink_sec}
Modern cosmological simulations can potentially use billions of particles
to model the formation of the Universe. However, it is still challenging 
to resolve mass scales from galaxy clusters (10$^{13}\,\Msun$) to a
stellar scale ($1\,\Msun$). For example, the resolution length in our 
simulation is about $1\,\pc$, hence modeling the process of star formation on cosmological scales from first 
principles is impractical for the current setup. Alternatively, in the treatment of star 
formation and its feedback, sub-grid models are employed, meaning that a 
single particle behaves as a star, which comes from the results of stellar models. 
Also, when the gas density inside the simulations becomes increasingly high, 
the SPH smoothing length decreases according to the Courant condition and forces it to shrink the time steps very rapidly. When the resulting runaway collapse occurs, 
the simulation easily fails. Creating sink particles is required to bypass this numerical 
constraint and to continue following the evolution of the overall system for longer. 
For the treatment of star formation, we apply the sink particle algorithm\cite{JB2007}{}.
We have to ensure that only gravitationally bound particles can be merged 
to form a sink particle and utilize the nature of the Jeans instability. 
We also consider how the density evolves with time inside the collapsing 
region of the first star formation when gas densities are close to $n_c\,\sim\,10^4\,\cm^{-3}$ and subsequently increase rapidly by several orders 
of magnitude. So the most important criterion for a particle 
to be eligible for merging is $n\,>\,n_{c}$ because in the collapse around the 
sink particles, the velocity field surrounded by the sink must be converged fluid,
which yields $\nabla\cdot\vec{v}<0$.  The neighboring particles around the sink 
particle should be bounded and follow with [\refcite{JB2007}]
\begin{equation}
 E\,=\,E_{g}\,+\,E_{k}\,+\,E_{t}\,<\,0,
\end{equation}
where $E$, $E_{g}$, $E_{k}$, and $E_{t}$ are the overall binding, gravitational, kinetic, and thermal 
energies, respectively. Sink particles are usually assumed to be collisionless, so that they only 
interact with other particles through gravity. Once the sink particles are 
formed, the radiative feedback from the star particles would halt further accretion of in-falling gas.  
So collisionless properties of sink particles are reasonable for our study. The sink particles 
provide markers for the position of a Pop~III star and its remnants, such as a black hole or supernovae, 
to which the detailed physics can be supplied.

\subsection{Radiative Transfer}
\label{ray_sec}
When a Pop~III star has formed inside the minihalo, the sink particle
immediately turns into a point source of ionizing photons to mimic the
birth of a star. The rate of ionizing photons emitted depends on the
physical size of the star and its surface temperature based on the subgrid 
models of stars. Instead of simply assuming constant rates of emission, 
we use the results of one-dimensional stellar evolution 
to construct the luminosity history of the Pop~III stars that served as our
sub-grid models for star particles. The luminosity of the
star is actually evolving with time and demonstrates a considerable
change. The streaming photons from the star then form an ionization front and build 
up \HII{} regions. For tracing the propagation of photons and the ionization front, 
we use the ray-tracing algorithm from [\refcite{greif2009a}], which solves the
ionization front equation in a spherical grid by tracking $10^5$ rays
with 500 logarithmically spaced radial bins around the ray source.
The propagation of the ray is coupled to the hydrodynamics of the gas through 
its chemical and thermal evolution. The transfer of the $\rm H_2$-dissociating photons
of Lyman--Werner (LW) band ($11.2-13.6\,\eV$) from Pop~III stars is also 
included.

In the ray-tracing calculation, the particles' positions are transformed from Cartesian to 
spherical coordinates, radius ($r$), zenith angle ($\theta$), and azimuth angle ($\phi$). 
The volume of each particle is  $\sim h^3$, when $h$ is the 
smoothing length.  The corresponding sizes in spherical coordinates are $\Delta r\,=\,h$,
 $\Delta \theta\,=\,h/r$, and  $\Delta \phi\,=\,h/r\sin(\theta)$. Using spherical coordinates is 
for convenience in calculating the Str\"{o}mgren sphere around the star, 
\begin{equation}
n_n r^2_{\rm I}\frac{d r_{\rm I}}{dt}\,=\, \frac{\dot{N}_{\rm ion}}{4\pi}\,-\,\alpha_{\rm B}\int_0^{r_{\rm I}}n_en_+r^2dr, 
\label{rad1}
\end{equation}
where $r_{\rm I}$ is the position of the ionization front, $\dot{N}_{\rm ion}$ represents 
the number of ionizing photons emitted from the star per second, $\alpha_{\rm B}$ is
the case B recombination coefficient, and $n_n$, $n_e$, and $n_+$ are the number densities of neutral
particles, electrons, and positively charged ions, respectively. The recombination coefficient 
is assumed to be constant at temperatures around $2\times10^4\,\K$. The ionizing photons for 
\HI{} and \HeII{} emitted are
\begin{equation}
\dot{N}_{\rm ion}\,=\,\frac{\pi L_*}{\sigma T_{\rm eff}^4}\,\int_{\nu_{\rm min}}^{\infty}\frac{B_{\nu}}{h_{\rm P}\nu} {\rm d}\nu, 
\label{rad2}
\end{equation}
where $h_{\rm P}$ is the Planck's constant, $\sigma$ is the Boltzmann's constant,  $\sigma_{\nu}$ is the photo-ionization 
cross sections, and $\nu_{\rm min}$ is the minimum frequency for the ionization photons of 
\HI, \HeI, and  \HeII. 
By assuming the blackbody spectrum of a star $B_\nu$ of an effective temperature, $T_{\rm eff}$,
its flux can be written 
\begin{equation}
F_{\nu}\,=\,\frac{L_*}{4 \sigma T_{\rm eff}^4r^2}B_{\nu}. 
\end{equation}
The size of the \HII{} region is determined by solving Equation~(\ref{rad1}). 
The particles within the \HII{} regions now save information
about their distance from the star, which is used to calculate the 
ionization and heating rates,

\begin{equation}
k_{\rm ion}\,=\,\int_{\nu_{\rm min}}^{\infty}\frac{F_{\nu}\sigma_{\nu}}{h_{\rm P}\nu} {\rm d}\nu, \quad\quad
\Gamma\,=\,n_n\int_{\nu_{\rm min}}^{\infty}F_{\nu}\sigma_{\nu}\Big(1\,-\,\frac{\nu_{\rm min}}{\nu}\Big){\rm d}\nu. \
\end{equation}
\Hm{} is the most important coolant for cooling the 
primordial gas, which leads to formation of the first stars. However, its hydrogen bond is weak 
and can be easily broken by photons in the LW 
bands between 11.2 and 13.6 eV. The small $\Hm$ fraction in the IGM 
creates only a little optical depth for LW photons, allowing them to propagate over a much
larger distance than ionizing photons. In our algorithm, self-shielding of H$_2$ is not 
included because it is only important when H$_2$ column densities are high. 
Here we treat the photodissociation of $\Hm$ in the optically thin limit and the dissociation rate in a 
volume constrained by causality  within a radius, $r =ct $. The dissociation rate 
is given by $k_{\rm H_2}=1.1\times10^8 F_{\rm LW}\,\usec^{-1}$, where $F_{\rm LW}$
is the flux within LW bands. 

\subsection{Supernova Explosion and Metal Diffusion}
\label{sn_sec}
After several million years, the massive Pop~III
stars eventually burn out their fuel, and most of
them die as supernovae. As we discussed in Part~I, 
the first supernovae are very powerful explosions 
accompanied by huge energetics and metals. 
In this subsection, we discuss how we model the SNe explosion in 
our cosmological simulation. 

When the star reaches the end of its lifetime, we remove the 
star particles from the simulation and set up the explosions 
by injecting the explosion energy to desired particles surrounded 
by the previous sink. Because the resolution of the simulation 
is about 1~pc, we cannot resolve the individual 
SNe in both mass and space. Here we assume the SN ejecta is 
disturbed around a region of 10~pc, embedding the progenitor 
stars, in which most kinetic energy and thermal energy of ejecta 
are still conservative. We attach the metals to these particles based on
the yield of our Pop~III SN model. The explosion energy of hypernovae 
and pair-instability SNe can be up to $10^{52}-10^{53}\,\erg$. 
For the iron-core collapse SN, it is about $1.2 \times 10^{51}\,\erg$. 

In our \GADGET{} simulations, we are unable to resolve the stellar scale below 
1~pc. However, the fluid instabilities of SN ejecta develop initially at a scale far below 1~pc. 
These fluid instabilities would lead to a mix of SN metals with the primordial IGM.
Therefore, mixing plays a crucial role in transporting the metal, which could be the most 
important coolant for later star formation. To model the transport of metals, 
we apply a SPH diffusion scheme\cite{greif2009b} based on the idea of turbulent
diffusion, linking the diffusion of a pollutant to the local physical
conditions. This provides an alternative to spatially resolving mixing
during the formation of supernova remnants. 

A precise treatment of the mixing of metals in cosmological simulations 
is not available so far because the turbulent motions responsible for 
mixing can cascade down to very small scales, far beyond the resolutions 
we can achieve now. Because of the Lagrangian nature of SPH simulations, it is 
much more difficult than the direct modeling of mixing by resolving the fluid 
instabilities in SPH than in grid-based codes. However, we can assume the motion 
of a fluid element inside a homogeneously and isotropically turbulent velocity 
field, such as a diffusion process, which can be described by
\begin{equation}
\frac{{\rm d}c}{{\rm d}t }\,=\,\frac{1}{\rho}\nabla\cdot(D\nabla c),
\end{equation}
where $c$ is the concentration of a metal-enriched fluid-per-unit mass; {\it D}
is the diffusion coefficient, which can vary with space and time;
and $\frac{\rm d}{\rm dt}$ is the Lagrangian derivative. 

After the SN explosion, metal cooling must be considered in the cooling network. We assume that C, O, 
and Si are produced with solar relative abundances, which are the dominant coolants for the first SNe. 
There are two distinct temperature regimes for these species. In low temperature regimes, 
 $T\,<\,2\,\times\,10^4\,\K$, we use a chemical network presented in [\refcite{glover2007}], which follows 
 the chemistry of C, C$^+$, O, O$^+$, Si, Si$^+$, and Si$^{++}$, supplemental to the primordial species discussed above. 
 This network 
 also considers effects of the fine structure cooling of C, C$^+$, O, Si, and Si$^+$. The effects of molecular 
 cooling are not taken into account. In high temperatures, 
$T\,\geq\,2\,\times\,10^4\,\K$, due to the increasing number of ionization states, a full non-equilibrium 
treatment of metal chemistry becomes very complicated and computationally expensive. Instead of directly 
solving the cooling network, we use the cooling rate table\cite{suther1993}{}, which gives effective cooling 
rates (hydrogen and helium line cooling, and {\it bremsstrahlung}) at different metallicities.  Dust cooling
is not included because the nature of the dust produced by Pop~III SNe is still poorly understood.

\subsection{Parallel Performance of {\GADGET} }
\label{gadget_sec}
\GADGET{} simulations that include several physical processes are very computationally expensive
and must be run on supercomputers. It is good to know the scaling performance of the code so that
we can better manage our jobs. To understand the parallel efficiency of \GADGET, we perform a
strong scaling study. The test problem is a $\Lambda$CDM problem including gas hydrodynamics of 
gas particles coupled with gravity of CDM, which started with the condition at $z = 100$ in a periodic 
box of linear size of 1~Mpc (comoving), using $\Lambda$CDM cosmological parameters with matter 
density $\Omega_m=0.3$, baryon density $\Omega_b=0.04$, Hubble constant $H_0 = 70\,\km\,\sec^{-1} \Mpc^{-1}$, 
spectral index $n_{\rm s}=1.0$, and normalization  $\sigma_8=0.9$, based on the CMB measurement from WMAP 
\cite{komatsu2009}{}. The total number of particles for this problem is about 80 million (40 million for gas and 40 million for dark matter). 
 This is the identical setup for our real problem, including the cooling and the chemistry of the primordial gas. 

\begin{figure}[ht]
\begin{center}
\includegraphics[width=\columnwidth]{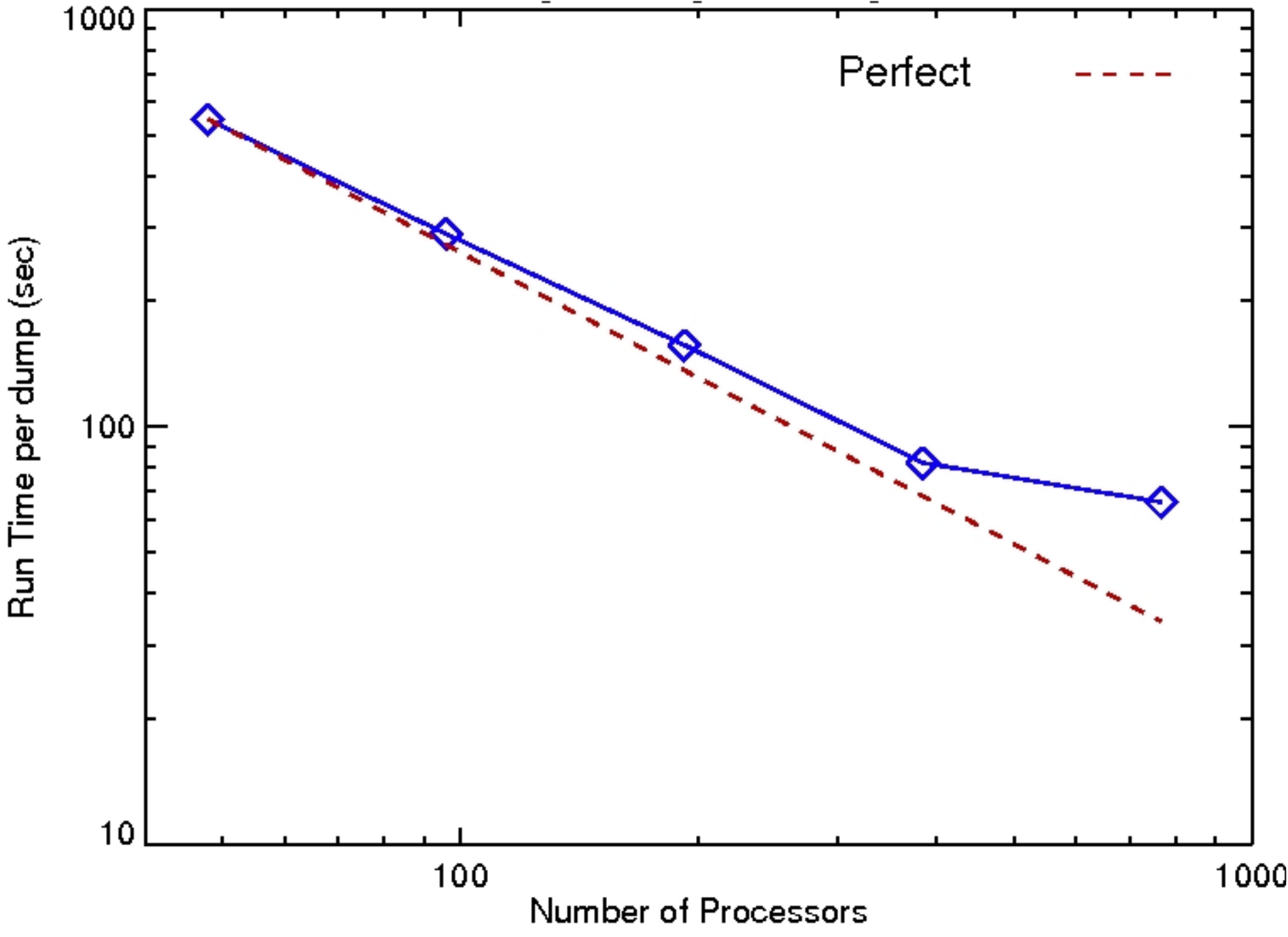}
\caption[]{Strong scaling of \GADGET{} on Itasca:
The blue curve presents the scaling performance of \GADGET{} on
Itasca, and the red-dashed curve is the case of perfect scaling. \label{gscaling}}
\end{center}
\end{figure}

The purpose of the scaling test is to allow us to determine the optimal computational resources to perform our simulations
and complete them within a reasonable time frame.  We perform these tests on Itasca, a 10,000$-$CPU supercomputer located at the Minnesota
Supercomputing Institute. We increase the CPU number while running the same job and record the amount of time it takes to finish 
the run. For perfect scaling, the run time should be inversely proportional to the number of CPUs used. Figure~\ref{gscaling}
presents the results of our scaling tests. It shows a good strong scaling when the number of CPUs is $n_c\,\lesssim\, 300$.  
Once $n_c\,>\,300$, the scaling curve becomes flat, which means the scaling is getting saturated, and $n_c\,=\,256$ seems 
to be a turning point. Hence we use $n_c\,=\,256-384$ for our production runs. 

\section{Pop~III Supernovae - Impact to the Early Universe}
\label{feed_result}
Galaxies are the building blocks of large-scale structures in the Universe. 
The detection of galaxies at $z \approx 10$ by the Hubble Space Telescope 
suggests that these galaxies formed within a  few hundred million years (Myr) 
after the Big Bang. In \S~\ref{firststar}, we discussed the Pop~III stars that 
are predicted to form inside the dark matter halos of mass about $10^5\,\Msun$, known as minihalos. 
The gravitational wells of minihalos are very shallow, so they could 
not maintain a self-regulated star formation because the stellar feedback from the Pop~III 
stars inside the minihalos could easily strip out the gas and prevent formation of the next 
subsequent stars. Thus the minihalos cannot be treated as the first galaxies. Instead, the first galaxies must 
be hosted by more massive halos generated from the merging of minihalos.  The high 
redshift galaxies should come from the merger of the first galaxies. But {\it how did the first galaxies form}? 
and {\it what are the connections among the first stars, the first supernovae, and the first galaxies?}

\begin{figure}[ht]
\begin{center}
\includegraphics[width = \columnwidth]{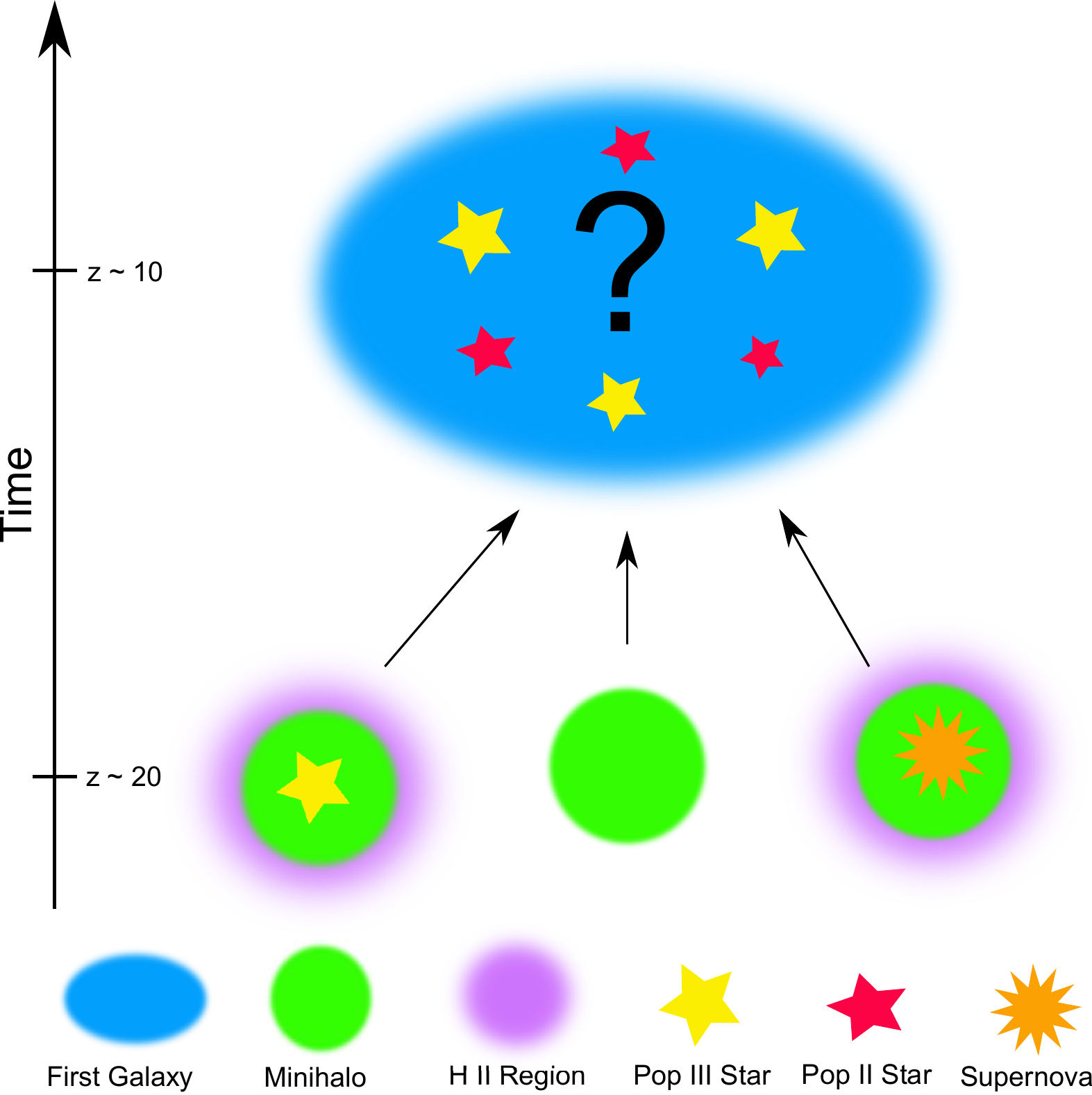}
\caption[]{ Assembly of the first galaxies: 
Based on the model of [\refcite{bromm2011}], the first galaxies form with a mass of about $10^{8}$ \Msun{} at $z \sim 10$. The feedback of previous Pop~III stars can affect the star formation inside the first galaxy. The gravitational wells of minihalos are shallow, so they cannot be treated as galaxies in this scenario. \label{afsg}}
\end{center}
\end{figure}

A key to answering these questions is held by the Pop~III stars formed inside the minihalos.  
Massive Pop~III stars might have died as supernovae (Pop~III SNe).  
The Pop~III stars with initial masses of $10-150\,\Msun$ die as CCSNe; 
those with initial masses of $150-260\,\Msun$ die as PSNe, and those with mass  
$ > 260\, \Msun$ just collapse to black holes.  We temporarily neglect the feedback of exploding super massive stars here 
because of their scarcity. Massive Pop~III stars could emit copious amounts of hydrogen-ionizing photons, which contribute to 
cosmic reionization.  Their SNe dispersed the first metals to the intergalactic medium (IGM). 
This chemical enrichment could trigger the formation of the second generation of stars (Pop~II stars). 
Finally, the minihalos and IGM, together with relic \HII{} regions and metals from Pop~III stars, jointly formed 
into the first galaxies, as shown in Figure~\ref{afsg}.

The formation of the first galaxies not only depends on the evolution of dark matter but also on baryon, 
which provides the material for forming stars. The chemical, mechanical, and radiative feedback from the first 
stars makes the assembly process of the first galaxies much more complex. The model of first galaxy formation 
is still at its infant phase and is not  sophisticated enough to offer reliable predictions. One of the obstacles for models 
is in resolving the relevant spatial scales and physical processes. Beneficial to the advancement of computational technology, 
new supercomputers allow us to perform more realistic cosmological simulations and start to investigate the first galaxy 
formation.

In this section, we review the current understanding of the first galaxies in \S~\ref{fg_sec}. 
Then we discuss the role of the first stars in the first galaxy formation in \S~\ref{ccstar_sec}. 
The stellar feedback includes radiation during its stellar evolution and chemical enrichment when
the star dies as a SN. We discuss the radiation feedback of the first stars in \S~\ref{fg_rad_sec} 
and the chemical enrichment of  their SNe in \S~\ref{fg_met_sec}.

\subsection{Assembly of the First Galaxies}
\label{fg_sec}
There are several definitions of the first galaxy. In general, a galaxy should have multiple stars hosted in a bound
halo; its potential well is deep enough to retain the gas heated by the UV radiation from stars or inside it
\cite{barkana2001,bromm2011,goodstein2011}. In addition, SN explosions in the first galaxies can only trigger 
a minimum mass loss. In brief, a galaxy must have a stable and self-regulated star formation.  
The potential well of the halo is the most important factor determining whether it can be a galaxy or not. For a given halo
mass  at $z \gg 1$, the gravitational binding energy of the halo can be estimated as [\refcite{bromm2011}],
\begin{equation}
 E_{b} = \frac{GM^2}{r_{\rm vir}} \simeq 5 \times 10^{53}\bigg(\frac{M}{10^8\,\Msun}\bigg)^{5/3} \bigg(\frac{\delta_c}{18\,\pi^2}\bigg)
 ^{1/3}\bigg(\frac{1+z}{10}\bigg)\,{\rm erg},
\end{equation} 
where $r_{\rm vir}$ is the virial radius of the halo, and $\delta_c$ is the density contrast when the halo formed. 
The results of [\refcite{wise2008,grief2010}] have suggested that dark matter halos of a mass of $10^8\,\Msun$ forming at $z\approx10$ can satisfy the 
criteria. These halos have a virial temperature of about $10^4\,\K$, which is related to the
characteristic temperature due to atomic hydrogen cooling.  These halos are also called atomic cooling halos. Unlike minihalos, 
the dominating cooling process of gas is by H instead of H$_2$. Such halos also keep most of their gas that previously 
received stellar feedback, such as through radiation and the SN blast wave. 
For observers, there are two primordial types of galaxies 
that can be the first galaxies. The first galaxies can be defined as the highest redshift galaxies detected. However, such 
a definition may change once there is a new telescope. On the other hand,  the galaxies containing zero metallicity may
be defined as the first galaxies.  However,  chemical enrichment might already occur in the first galaxies. 
In this review, we use definitions based on [\refcite{bromm2011}] for the first galaxies that are constructed by 
a dark matter halo and host the Pop~III or Pop~II stars.

\subsection{Cosmological Impact of the First Stars}
\label{ccstar_sec}
The process of the first galaxy formation is highly complex because the initial conditions and relevant physics 
are not well understood.  In the $\Lambda$CDM model, the first stars are predicted to have been born before the 
first galaxies formed. Thus the first stars together with primordial gas would offer a rockbed for the first galaxies. 
Feedback from the first stars would play an important role in determining the initial conditions for forming the first 
galaxies. The stellar feedback usually includes radiative \cite{schaerer2002}{} and supernova feedback \cite{ciardi2005}{}. 
The massive Pop~III stars produce UV radiation to ionize the primordial gas \cite{barkana2007}{}. The 
WMAP measured  an increasing optical depth at  $z\sim15$, implying cosmic reionization by the massive Pop~III stars. The 
SN feedback has both a mechanical and a chemical impact; the blast wave of the explosion injects heat and momentum to the 
surrounding IGM and concurrently disperses metals into the primordial gas [\refcite{ferra2000,bromm2004a}].
As discussed before, some Pop~III stars may die as PSNe, and such explosion modes could quickly pollute the IGM 
with large amounts of metals. Such chemical enrichment can alter the subsequent star formation because additional 
metal cooling starts to function. Both radiative and SN feedback of the first stars transforms the simple Universe into a much 
more complex state by setting the initial conditions for the first galaxy formation. 

\begin{figure}[ht]
\begin{center} 
 \includegraphics[width=\columnwidth]{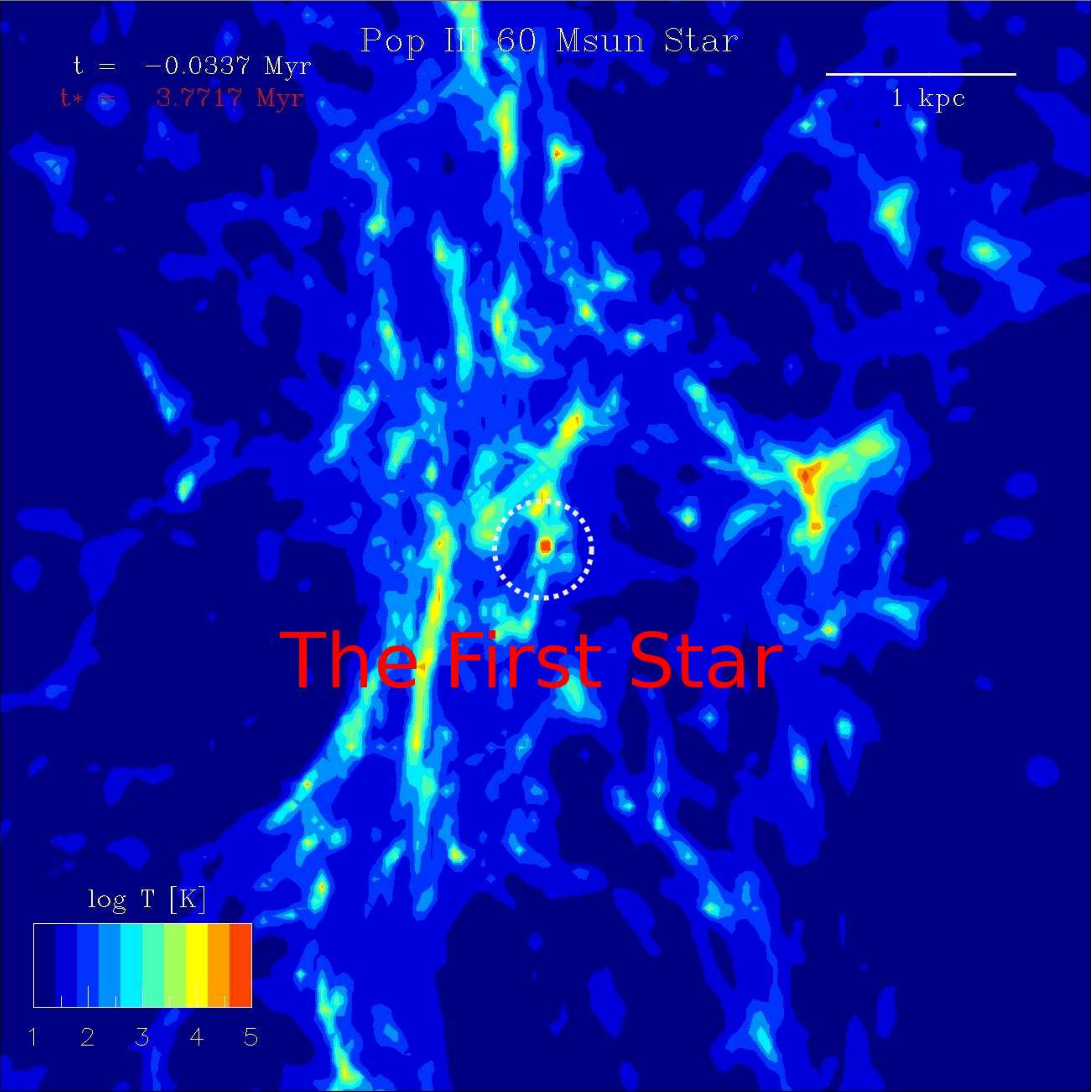}
\caption[]{ Birth of the first stars: The first star is about to form within the dark matter 
halo (white circle) of mass  of $10^5\,\Msun$ at $z\approx28$. There will be a runaway collapse, and a sink will form to mimic the star formation. \label{bsg}}
\end{center}
\end{figure}

Figure~\ref{bsg} shows a density snapshot of our cosmological simulations at 
the time when the first star is about to form inside one of the minihalos. The density of the gas
cloud is approaching $10^4\,\cm^{-3}$, and its \Hm{} mass fraction rises to $10^{-3}$. 
to cool the gas cloud to about $200\,\K$.  A runaway collapse of the cloud will occur, and
the first star is about to form.

\subsection{Radiative Feedback}
\label{fg_rad_sec}
The radiation emitted from Pop~III stars would affect the subsequent
thermal properties of the primordial IGM, which changed the properties 
of star-forming clouds and affected the later star formation inside the first 
galaxies. The radiative feedback may have several
different forms, e.g., UV photons and x-rays, depending on the stars
and their compact remnants. Since H$_2$ is the most important coolant for 
the first star formation, it is relevant to learn how the radiation 
influences  H$_2$.  The hydrogen bond of H$_2$ is weak and can be easily broken 
by Lyman--Werner (LW) photons with energy in $11.2-13.6$~eV, 
\begin{equation}
 {\rm H}_2\,+\,\gamma\,\rightarrow\,{\rm H}^*_2\,\rightarrow\,2{\rm H}.
\end{equation}
H$_2^*$ is an excited state, which is unstable and soon decays into two H.
Massive Pop~III stars could emit large amounts of UV photons, 
easily ionizing the primordial hydrogen and helium, thus suppressing the 
corresponding H$_2$ cooling. Without effective H$_2$ cooling, massive 
Pop~III stars may not be able to form hereafter. On the other hand, in the ionized 
region, the abundance of free electrons may increase and facilitate the formation of H$_2$.
It is still unclear whether the radiation from the Pop~III stars is helpful (facilitating later star formation) 
or harmful (hampering later star formation). The overall impact of the radiative feedback on 
the H$_2$  is pretty uncertain. Besides ionizing primordial gas, energetic UV photons can photoheat 
the surrounding gas and allow it to escape the host halo and form an outflow. This disperses 
the gas inside the minihalos and may shut the later star formation off. More cosmological simulations of 
comprehensive radiative effects of the Pop~III stars are necessary for clarifying this issue. 
 
\begin{figure}[ht]
 \begin{center}
 \includegraphics[width=\columnwidth]{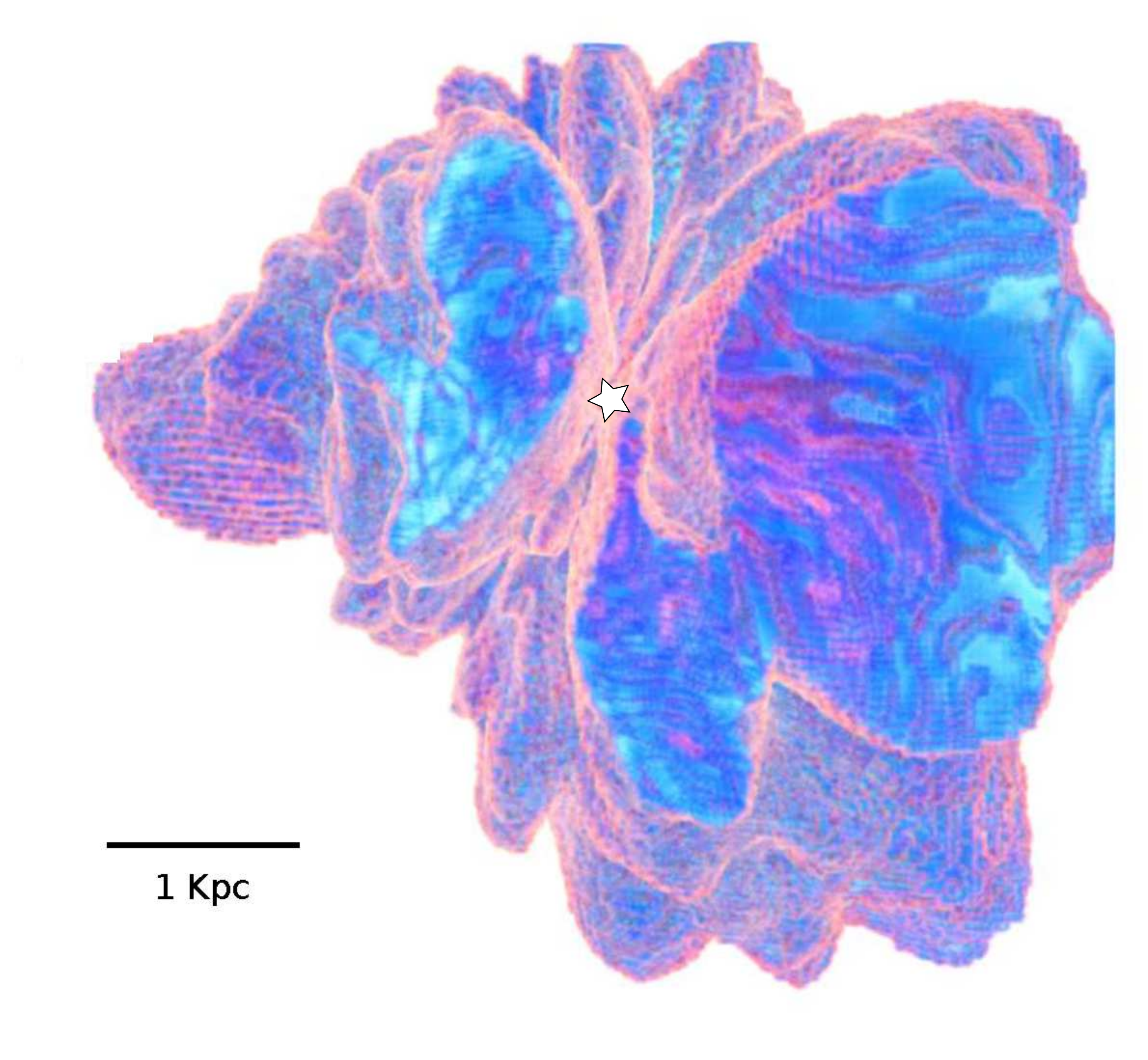}
 \caption[]{ The \HeII{} region created by a $100\,\Msun$ Pop~III 
 star before it dies: The white star indicates the position of the star. The strong UV 
 photons emitted from the star create an extensive \HeII{} region of a size about several $\kpc$. \label{ionhep}}
 \end{center}
 \end{figure}

 Figure~\ref{ionhep} shows a \HeII{} region created by a $100\,\Msun$ Pop~III star.
 When the first star evolves to the main sequence and stable hydrogen burning at the core occurs, 
 its surface temperature quickly rises to $T\sim2\times10^5$ K and begins to emit a large amount of 
 ionizing photons for neutral hydrogen and helium. The gas inside the host halo is strongly 
 photoheated, which allows the gas to escape the
 gravitational well of the host halo, forming an outflow.
 
 \subsection{Supernova Feedback}
\label{fg_met_sec}
Massive Pop~III stars might die as energetic SNe and dump metal-rich ejecta to the IGM. 
The are two kinds of feedback from SNe: thermodynamical and chemical. 
The SN explosions produce strong shocks that blow up the stars (see Part~I).
The SN feedback strongly depends on its progenitor stars, which determine the amount of 
explosion energy and metals produced.  Chemical enrichment of the IGM by Pop~III 
SNe is important for understanding the transition in the star-formation mode from high-mass dominated 
to low-mass dominated [\refcite{bromm2004a}]. If metals are very uniformly dispersed by the SNe, the transition
may  occur rather sharply.  In contrast, if the enrichment is not very uniform, gas clumps of high-metallicity 
may appear and surround the primordial gas. In this case, the transition of star formation mode 
may occur more smoothly. [\refcite{bromm2003}] first  present numerical simulations of the first SN explosions at 
high redshifts ($z \approx 20$); they assume that one single PSN occurs inside the center of the minihalo 
and simulate the explosion. Their simulations explore two explosion energies of PSNe, 
$10^{51}\,\erg$ and  $10^{53}\,\erg$. Their results show that  the explosion of $10^{53}\,\erg$ can 
create giant metal bubbles the size of several kpc. The lower explosion energy instead shows relatively smaller 
regions of metal enrichment.  More recent results \cite{grief2010} show that the metals are dispersed
uniformly due to the diffusion mixing. 

\begin{figure}[ht]
\begin{center}
\includegraphics[width=\columnwidth]{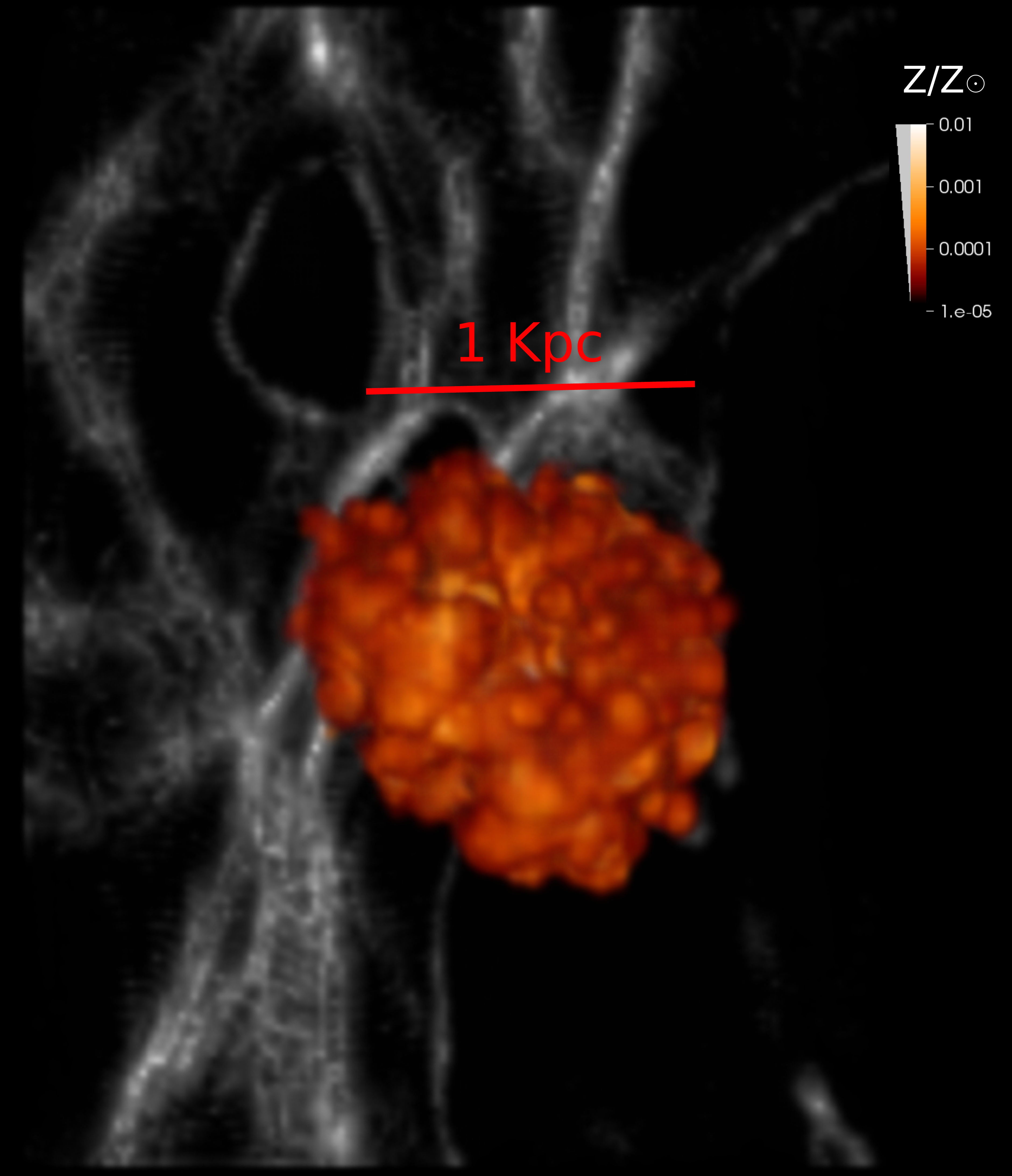}
\caption[]{ Metal enrichment of the first SNe: The \GADGET{} simulation shows that a SN explosion from a $60\,\Msun$  star can efficiently spread the metal over 1~kpc in a 
few million years and enrich the metallicity of pristine gas inside IGM to $10^{-3}$ - $10^{-5}\Zsun$.  \label{fsgsn}}
\end{center}
\end{figure}

Figure~\ref{fsgsn} shows a SN explosion at five million years after its onset. The metal of the SN has been 
dispersed to the IGM of a radius about $1\,\kpc$. 
When the SN shock breaks out of the stellar surface and propagates into the low-density 
ISM surrounding, it is suddenly accelerated to a velocity above $10^4$ km/s, about a few 
percent of the speed of light. The shock front can quickly reheat the relic \HII{} regions 
created by the progenitor stars and 
maintain the ionized status of the \HII{} region for 
an additional $1\sim2\,\Myr$. For chemical feedback, the SN ejecta are metal-rich and 
can pollute the pristine IGM to a metallicity of about $10^{-3}$-$\,10^{-5}\,\Zsun$  inside 
a region of 1~kpc. The first metals are very important to the later star formation because the 
metal cooling affects the mass of scale during the star formation. Once the gas cloud reaches the critical metallicity\cite{schneider2012}{}, $10^{-3}\,\Zsun$, Pop~II stars that 
have a mass scale similar to present-day stars may start to form. 
The resolutions of these simulations are still very crude. We are just starting to understand the complex processes of the first chemical enrichment by the first SNe.

\section{Summary and Perspective}
\label{summary}
One of the frontiers in modern cosmology is understanding the end of the cosmic dark ages, 
when the first luminous objects (e.g., stars, supernovae, and galaxies) transformed the 
simple early Universe into a state of ever-increasing complexity. In this review,
we discussed several possible fates of the first SNe as well as their impact on the 
early Universe. 

The thermonuclear supernovae of very massive stars include two types of
pair-creation instability supernovae and one possible type of general relativity supernovae.
The first stars with initial masses of $80-150\,\Msun$ might eject a few solar masses pulsationally; 
they are triggered by violent instabilities in stellar cores before they die. These ejected masses may lead to catastrophic collisions and power extremely 
luminous optical transients called pulsational pair-instability supernovae, which may account for the 
superluminous supernovae. The first stars with initial masses of $150-260\,\Msun$ eventually die as pair-instability supernovae.  We report the discovery of an extraordinary supernova of a 
$55,500\,\Msun$. 
We infer that the possible driver of the explosion of a super massive star is triggered by
general relativity, where the supporting pressure term becomes a source of gravity. 
This catalyzes the helium burning, leading to an explosion of energy up to $10^{55}\,\erg$, which is 
about 10,000 times more energetic than normal SNe. This also implies a narrow mass window 
in which the super massive stars may die as supernovae instead of collapsing into black holes. 
Violent mixing has been found inside the GSNe ejecta. 
These SNe produce a broad range of fluid instabilities and resulting mixing that is 
reflected in their observational signatures.

We discuss the impact of the first stars and their SNe on the early Universe.
The stellar feedback from the first stars could affect the later star formation 
and the assembly of the first galaxies.  Because the proper mass scale of 
the first stars and their population are very uncertain. The stellar impact depends on the mass of the stars; 
the more massive the stars are, the more UV photons can be produced, which leads to a 
more extensive region of ionized hydrogen and helium. Massive Pop~III stars can die as several different 
kinds of supernovae, such as core-collapse supernovae and hypernovae, yielding different explosion 
energetics and amounts of metals. The metals dispersed by SNe can enrich the primordial gas and may 
lead to the formation of the second generation of stars forming inside the first galaxies. 
Our results suggest that the first stars of masses can effectively create a \HII{} region of a size about $3-4\,\kpc$ 
and enrich a region of IGM gas of size $1 - 2\,\kpc$ to a metallicity of $\sim 10^{-3}-10^{-5}\,\Zsun$. 
The chemical enrichment tends to be uniformly painted on the primordial gas instead of forming  higher-metallicity clumps. 

Simulations shed a light of understanding on the underlying physical processes 
of the first supernovae and their impacts. However, astronomy is a science 
based on observational data. Models only offer a promising way of understanding
the data. Strong theoretical models from simulations must offer useful 
predictions for observation, such as light curves or the spectra of targeted objects. 
For calculating predictions for these first SNe, a self-consistent radiation transport 
must be considered. 
Hydrodynamics simulations, including radiation calculations, can be very 
computationally expensive and technically difficult. One high-resolution 3D SN simulation 
may require several million CPU hours and can only be run on some of the world's most powerful 
supercomputers. Much effort are still needed to push the model frontiers.

The first supernovae hold the keys to understanding how the cosmic dark ages were terminated. 
The detection of these objects will be the holy grail in modern cosmology. New ground and space 
telescopes with unprecedented apertures are planned for achieving this goal (see Table~\ref{ta_telescopes}). These forthcoming ground-based facilities include 
the {\sf European Extremely Large Telescope (E-ELT)}, the {\sf Thirty Meter Telescope (TMT)}, 
and the {\sf Giant Magellan Telescope (GMT)}. In space, the {\sf James Webb Space Telescope} 
({\sf JWST}) will take the lead. 

These telescopes will become the world's biggest eyes in the sky in human history and will allow us 
to probe the most distant Universe, showing when the first luminous objects such as stars, supernovae, and galaxies 
were about to form. Meaningful predictions of the first luminous objects through robust simulations 
are critical to the success of these observatories, which will be constructed by 2020. Before that date, 
significant efforts are needed to refine models to achieve the level of sophistication that will 
offer the most accurate scientific predictions for these forthcoming facilities. It is 
extremely urgent and important that we start to push the model frontiers along with the construction 
of these telescopes. With fast-growing computational power, simulations will be able to resolve the spatial scale as well as 
relevant physical processes that occur.  
With both the forthcoming data and the sophisticated models, the most enigmatic and radical mystery of 
these first luminous objects will be revealed in the foreseeable future.

  \begin{table}[ht]
   \tbl{Future telescopes for studying the early Universe}
   {\begin{tabular}{@{}ccccc@{}} \toprule
 Name  &  Type  & Aperture (m) &  Planned  &  References \\ \colrule
 {\sf E-ELT} & Ground &  40  &  2020+  & [\refcite{evans2013}] \\
 {\sf JWST} & Space & 6.5 &  2018+  & [\refcite{gardner2006}] \\ 
 {\sf TMT} & Ground & 30 &  2018+  & [\refcite{nelson2008}]\\
 {\sf GMT} & Ground & 24.5 &  2018+ & [\refcite{johns2012}] \\  \botrule
   \end{tabular} \label{ta_telescopes}}

   \end{table}

\section*{Acknowledgments} 
I thank Alex Heger, Stan Woosley, Volker Bromm, Ann Almgren, Lars Bildsten, John Bell, 
and Dan Kasen for many useful discussions.  K.C. was supported by an IAU-Gruber 
Fellowship, a Stanwood Johnston Fellowship, and a KITP Graduate Fellowship.  All numerical 
simulations were performed with allocations from the University of Minnesota Supercomputing 
Institute and the National Energy Research Scientific Computing Center.  This work has been 
supported by the DOE grants; DE-SC0010676, DE-AC02-05CH11231, DE-GF02-87ER40328, DE-FC02-09ER41618 and by the NSF grants; AST-1109394, and PHY02-16783.


\end{document}